\documentclass[12pt]{article}
\usepackage{amsmath}
\usepackage{graphicx}
\usepackage{enumerate}
\usepackage{natbib}
\usepackage{url} 
\usepackage{multirow}
\usepackage{float}
\usepackage{slashbox}
\usepackage{caption}
\usepackage{tikz}
\usepackage{amsthm,amsmath,amssymb}
\usepackage{diagbox}
\usepackage{subfigure}
\usepackage{booktabs}
\usepackage{diagbox}
\usepackage{caption}

\newcommand{\blind}{0}
\newtheorem{thm}{Theorem}

\allowdisplaybreaks[3]

\begin{document}

\def\spacingset#1{\renewcommand{\baselinestretch}%
{#1}\small\normalsize} \spacingset{1}


\if1\blind
{
	\bigskip
	\bigskip
	\bigskip
	\begin{center}
		{\LARGE\bf A Geometric Statistic for Quantifying Correlation Between Tree-Shaped Datasets}
	\end{center}
	\medskip
} \fi

\if0\blind
{
	\title{\bf A Geometric Statistic for Quantifying Correlation Between Tree-Shaped Datasets}
	\author{Shanjun Mao
		\hspace{.2cm}\\
		College of Finance and Statistics, Hunan University\\
		and \\
		Xiaodan Fan$^1$ \\
		Department of Statistics, The Chinese University of Hong Kong\\
		and \\
		Jie Hu$^1$ \\
		Department of Probability and Statistics, School of Mathematical Science,\\
		Xiamen University}
	\maketitle
	
	\footnotetext[1]{To whom correspondence should be addressed: xfan@cuhk.edu.hk, hujiechelsea@xmu.edu.cn}
} \fi

\bigskip
\begin{abstract}
The magnitude of Pearson correlation between two scalar random variables can be visually judged from the two-dimensional scatter plot of an independent and identically distributed sample drawn from the joint distribution of the two variables: the closer the points lie to a straight slanting line, the greater the correlation. To the best of our knowledge, similar graphical representation or geometric quantification of tree correlation does not exist in the literature although tree-shaped datasets are frequently encountered in various fields, such as academic genealogy tree and embryonic development tree. In this paper, we introduce a geometric statistic to both represent tree correlation intuitively and quantify its magnitude precisely. The theoretical properties of the geometric statistic are provided. Large-scale simulations based on various data distributions demonstrate that the geometric statistic is precise in measuring the tree correlation. Its real application on mathematical genealogy trees also demonstrated its usefulness.
\end{abstract}

\noindent%
\vfill

\newpage
\spacingset{1.9} 
\section{Introduction}
\label{sec:intro}

There are various tree-shaped datasets attracting increasing attention from many research and economic fields, such as gene expression data measured on a cell lineage tree \citep{hu2015bayesian}, spreading paths of information or infectious diseases \citep{ZhangThe}, and genealogy trees with parent-child relationships \citep{horowitz2008mathematicians}. To analyze such tree-shaped datasets, a fundamental issue is to quantify tree correlation, i.e., how the values from two trees of the same topology go up and down together along the same paths on the trees. \cite{mao2021tree} introduced a parameter similar to the Pearson correlation coefficient in the bivariate Gaussian distribution for quantifying the tree correlation, and provided a Bayesian approach to estimate the parameter. However, the tree correlation parameter by \cite{mao2021tree} lacks the appealing geometric interpretation as the Pearson correlation, i.e., the closeness of data points to a straight slanting line, which is the main concern to be addressed in this paper.

\begin{figure}[htb]
	\centering
	\includegraphics[scale=0.5]{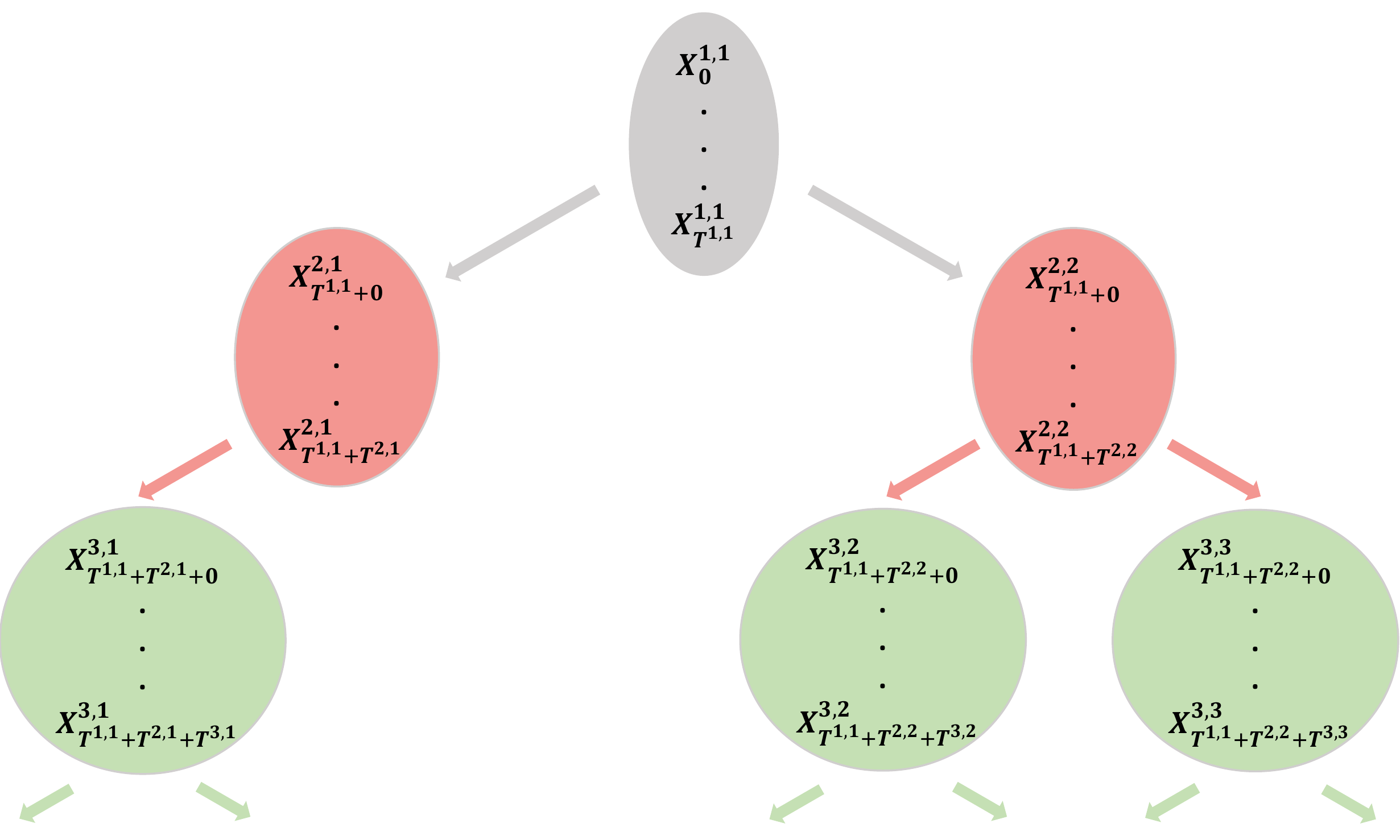}
	\caption{An example of tree-shaped dataset with binary tree topology. $X^{i,j}_{obs\_time}$ represents the observation of the $j$-th node in the $i$-th generation of the tree $\boldmath{X}$ at time $obs\_time$. Note that the tree depth may correspond to space instead of time. In both cases, $obs\_time$ is generally deemed as a position index.}
	\label{treedata}
\end{figure}

\begin{figure}[htb]
	\vspace{-0.8cm}
	\subfigure[]
	{\begin{minipage}{8cm}
			\centering
			\includegraphics[scale=0.55]{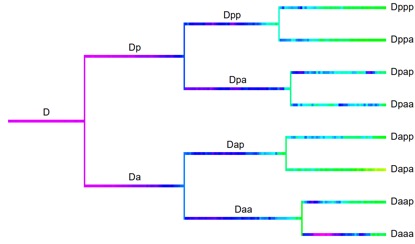}
	\end{minipage}}
	\subfigure[]
	{\begin{minipage}{8cm}
			\centering
			\includegraphics[scale=0.35]{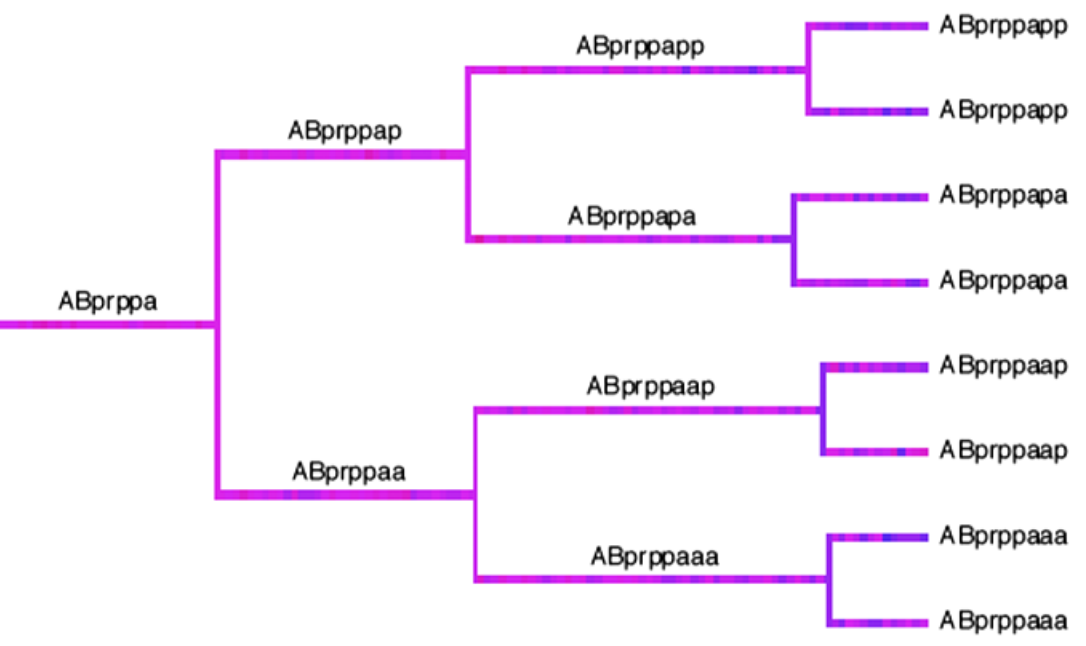}
	\end{minipage}}
	\caption{A real example of tree-shaped dataset. Each sub-figure displays the gene expression value of a specific \textit{C. elegans} gene along a cell lineage tree. Each node, represented by a horizontal line, denotes a cell with a name like `D', `Daa', etc. A horizontal line of changing colors represents the corresponding cell's time series observations. The vertical line connects a parent cell to its corresponding child cells. The color represents the measured gene expression values, which gradually increases with the color changing from purple to yellow according to the rainbow color order.}
	\label{realdataexample}
\end{figure}

The topological graph in Figure~\ref{treedata} illustrates the definition of tree-shaped data. In a tree, each node can have at most one parent node. Different nodes may have different numbers of observations. Nodes of the same tree depth form a generation. The root node is regarded as the first generation. Figure~\ref{realdataexample} shows an real example of tree-shaped dataset from \cite{hu2015bayesian}. In \cite{mao2021tree}, a Gaussian tree-shaped model was proposed as follows:
\begin{footnotesize}
	\begin{equation}
		\begin{split}
			& \begin{pmatrix}
				X^{i,j}_{T^{A(\cdot,i,j)}+0} \\ Y^{i,j}_{T^{A(\cdot,i,j)}+0}
			\end{pmatrix}
			=
			\begin{pmatrix}
				X^{A(i-1,i,j)}_{T^{A(\cdot,i,j)}} \\ Y^{A(i-1,i,j)}_{T^{A(\cdot,i,j)}}
			\end{pmatrix} \ ,\quad \text{for}\ i =2, 3, \cdots,\\
			& \left(\begin{array}{c|c}
				X^{i,j}_{T^{A(\cdot,i,j)}+t}& \\
				Y^{i,j}_{T^{A(\cdot,i,j)}+t}&
			\end{array} \begin{array}{cc}
				X^{i,j}_{T^{A(\cdot,i,j)}+t-1} = x^{i,j}_{T^{A(\cdot,i,j)}+t-1}& \\
				Y^{i,j}_{T^{A(\cdot,i,j)}+t-1} = y^{i,j}_{T^{A(\cdot,i,j)}+t-1}&
			\end{array} \right)
			\sim N
			\begin{pmatrix}
				\begin{pmatrix}
					\mu_i^x + x^{i,j}_{T^{A(\cdot,i,j)}+t-1} \\ \mu_i^y + y^{i,j}_{T^{A(\cdot,i,j)}+t-1}
				\end{pmatrix}
				&
				,
				&
				\begin{pmatrix}
					\sigma_1^2 & \rho^i \sigma_1 \sigma_2 \\
					\rho^i \sigma_1 \sigma_2 & \sigma_2^2
				\end{pmatrix}
			\end{pmatrix}, \\
			& \quad \text{ for } i = 1, 2, \cdots, \ \ \ t = 1, 2, \cdots, T^{i,j},
		\end{split}
		\label{gaussianmodel0}
	\end{equation}
\end{footnotesize}
where $T^{A(\cdot,i,j)} = \sum_{r=1}^{i-1} T^{A(r,i,j)}$, $A(r,i,j)$ is the index of the ancestral node of node $(i,j)$ in the $r$-th generation, and let $T^{A(\cdot,i,j)}=0$ when $i=1$. The first formula ensures the continuity of data by setting the initial observation of a node as a dummy variable and making it the same as the last observation of the corresponding parent node. The second formula shows the dependence within a node and the dependence of two trees simultaneously by a conditional distribution.

\begin{figure}[htb]
	\vspace{-0.8cm}
	\subfigure
	{\begin{minipage}{8cm}
			\centering
			\includegraphics[scale=0.3]{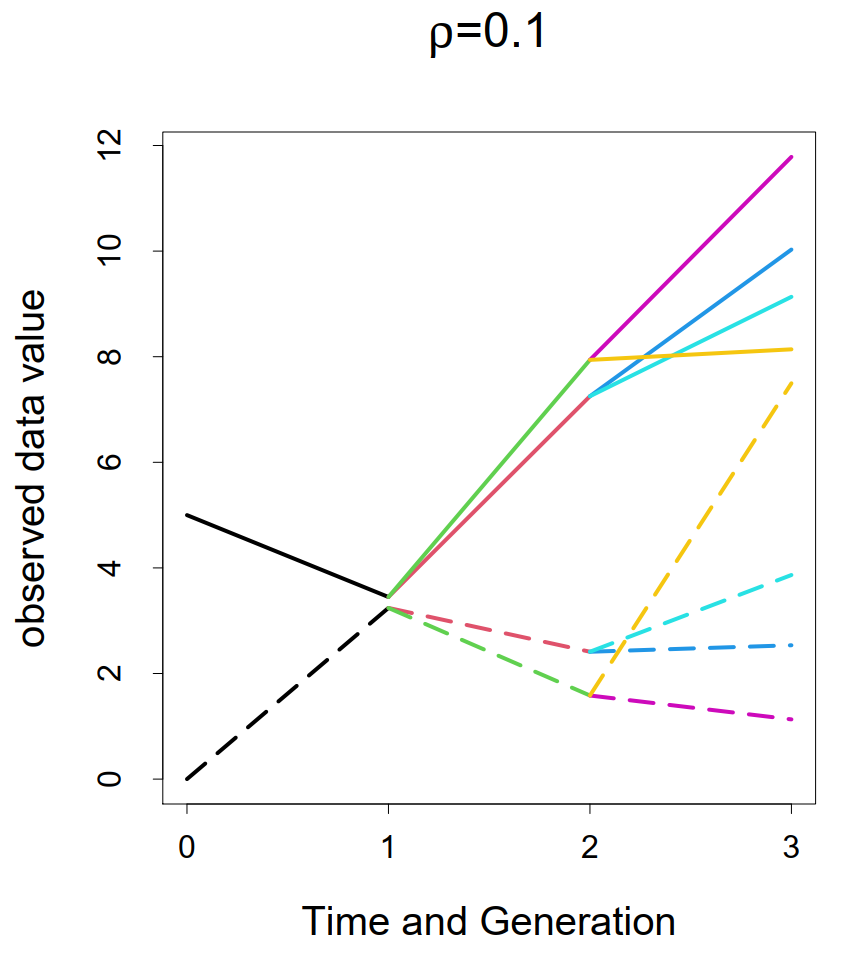}
	\end{minipage}}
	\subfigure
	{\begin{minipage}{5cm}
			\centering
			\includegraphics[scale=0.3]{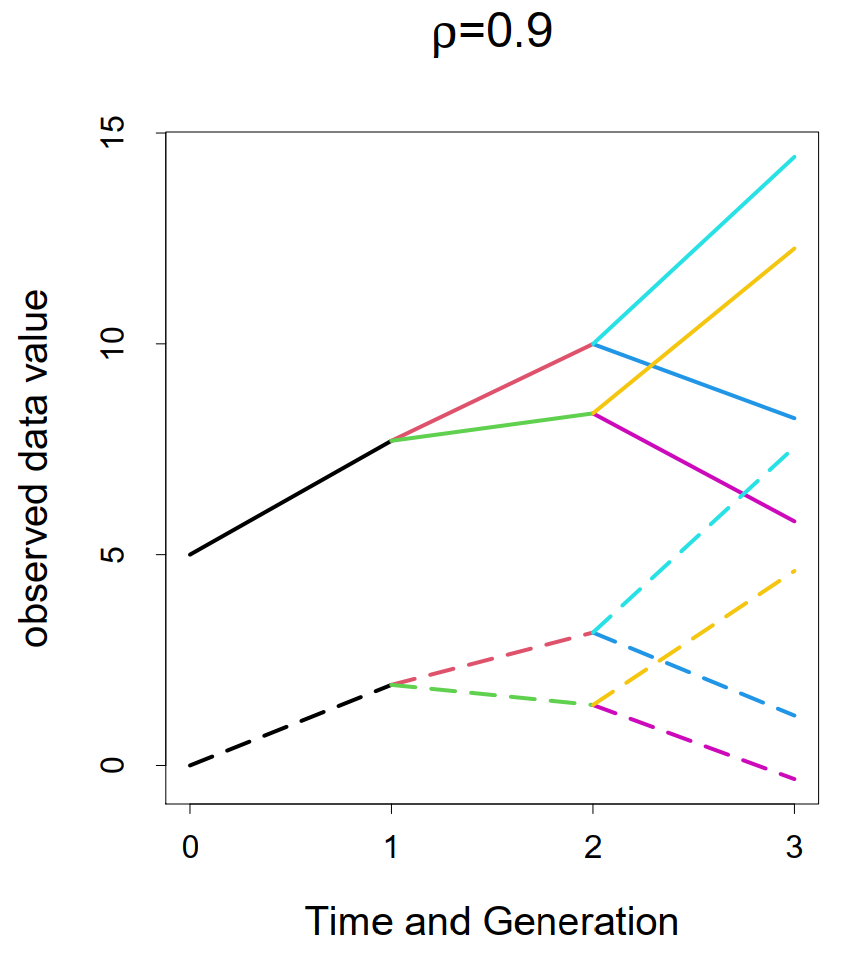}
	\end{minipage}}
	\caption{Synthesized tree-shaped datasets based on Model~\eqref{gaussianmodel0}. Here each node has one new observation, i.e., $T^{i,j} = 1$ for all $(i,j)$, $\mu_i^x = \mu_i^y = 2, \sigma_1^2 = \sigma_2^2 = 2$, and the corresponding $\rho$ is labeled on top of each sub-figure. In each sub-figure, the solid and dotted lines represent the observations from two different tree-shaped datasets $\boldsymbol{X}$ and $\boldsymbol{Y}$. The observations of two corresponding nodes from the tree pair are represented by the same color.}
	\label{tree_correlation_example}
\end{figure}

What we are interested in is the association of the data along the paths between two tree-shaped datasets. An example of tree-shaped datasets generated from Model~\eqref{gaussianmodel0} with $T^{i,j} = 1$ for each node $(i,j)$ is displayed in Figure~\ref{tree_correlation_example}. We can easily find that, the growth trend along each corresponding tree path is more synchronous when $\rho$ is larger. Hence, this kind of association between the tree pair can be interpreted as trend similarity or the degree of coupling. In our model settings, the trend similarity between two tree-shaped datasets is represented by synchronously increasing or decreasing at matched positions of the trees. At matched positions, the correlation magnitude of the corresponding variables may vary with the depth of the tree, such as exponential decay mechanism in Model~\eqref{gaussianmodel0}. In conclusion, we call this type of association between tree-shaped datasets as tree correlation.

In Model~\eqref{gaussianmodel0}, the parameter $\rho$ was regarded as the representation of tree correlation and Bayesian approach was applied to estimate the parameters \citep{mao2021tree}. However, the generalization ability of the method is constrained since the model is parametric. In addition, this method fails to show the tree correlation's geometric interpretation, which are helpful for more perceptive presentation and more vivid understanding \citep{vos2010geometry}. It is well known that some classical statistical methods are related to geometric interpretation and graphics. On one hand, many methods can be well explained from geometric perspective: Ordinary Least Square estimator can be understood as the projection from explained variable to the hyperplane expanded by explanatory variables, and Least Angle Regression \citep{Efron2004Least} algorithm can be regarded as an approach to solve Least Absolute Shrinkage and Selection Operator \citep{Robert1996Regression}. On the other hand, graphs help people to do statistical inference more conveniently and effectively by understanding and expressing meanings of statistics intuitively: quantile-quantile plot \citep{wilk1968Probability} helps to classify whether two datasets come from the same population, and Box plot \citep{Hintze1998Violin} exhibits both location and variation information of different groups of data. Therefore, graphic methods are effective and often adopted. For example, \cite{shamos1976geometry} developed new fast algorithms to compute familiar statistical quantities through a geometric viewpoint; \cite{umulis2012importance} proposed that the proper choice of geometry was critical for statistical or mathematical modeling of embryonic development. In particular, Pearson correlation coefficient \citep{galton1886regression} measures the linear correlation by showing how close the observations to an oblique line that is not parallel to the X and Y axes, as shown in Figure~\ref{pearsonplot}. This property promotes the wide application of Pearson correlation coefficient because we can just plot the observations to visually judge the correlation magnitude roughly.
\newsavebox{\smlmatdistributiona}
\savebox{\smlmatdistributiona}{$N \begin{pmatrix} \begin{pmatrix} 0 \\ 0 \end{pmatrix} &, & \begin{pmatrix} 1 & \rho \\ \rho & 1 \end{pmatrix} \end{pmatrix}$}
\begin{figure}[htb]
	\vspace{-0.2cm}
	\captionsetup{singlelinecheck=off}
	\centering
	\includegraphics[scale=0.25]{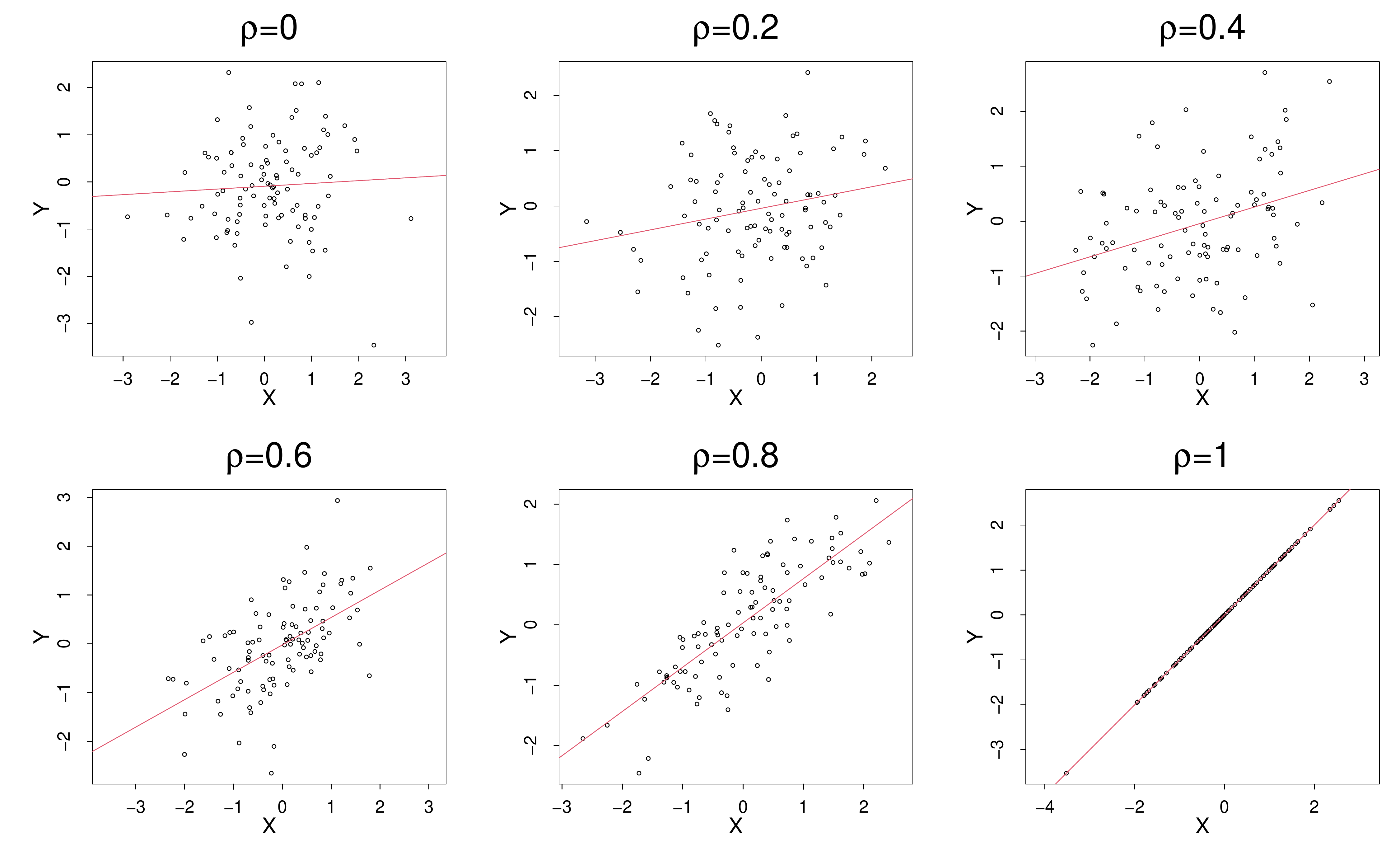}
	\caption{Geometric interpretation of Pearson correlation coefficient. For each diagram, 100 points are sampled from a bivariate Gaussian distribution \usebox{\smlmatdistributiona}, where $\rho$ is labeled at the top of each panel. The red line in each diagram represents the fitted regression line by Ordinary Least Square method.}
	\label{pearsonplot}
	\vspace{-0.7cm}
\end{figure}

However, for the tree correlation between two tree-shaped datasets, like in Figure~\ref{realdataexample}, we can find that, it may be hard to visually judge its correlation magnitude through a simple line as shown in Figure~\ref{pearsonplot}. Actually, if the observations of the same depth within the tree have the same correlation strength, a straight line can be used to reflect the correlation strength for each depth. But it is inappropriate to extend the method to the whole tree which has changeable correlation degree as tree depth. First, ignoring the relationships within correlations of different generations, i.e., the mechanism how correlation changes along the depth, will lead to a higher error. Second, the existence of tree topology not only leads to the imbalance of sample size in each generation, but more importantly, makes the data along paths across different branches dependent, like the self-correlation within time series. Thus, the classical correlation measures are not applicable. Our aim is to propose a new graphical method to both quantify the magnitude of tree correlation and quantify it.

In this paper, we provide an intuitive way to quantify tree correlation, more specifically, a geometric statistic which can be both visually judged from a special scatter plot of the data and calculated exactly by an algorithm. The geometric statistic can sort the degree of tree correlation precisely. Furthermore, if the data on the trees indeed follow a semi-parametric model with a Pearson correlation parameter, the geometric statistic is a monotone function of the Pearson correlation parameter. The remaining content of the paper is organized as follows: a preliminary model-free approach is introduced in Section~\ref{sec:pre}, a semi-parametric Gaussian model are proposed in Section~\ref{sec:model}; a geometric interpretation method of tree correlation, including theoretical proof and normalization algorithm, are presented in Section~\ref{sec:graph};  main results of large-scale simulation studies and real data analysis are demonstrated in Section~\ref{sec:simul} and Section~\ref{sec:real}, respectively; Section~\ref{sec:conclusion} gives some discussions and concludes the paper.

\section{Preliminary Approach}
\label{sec:pre}

Without loss of generality, Gaussian and Gamma distributions are adopted to generate example datasets in order to introduce the preliminary approach of quantifying tree correlation geometrically. Two kinds of synthesized tree-shaped datasets whose increments are following bivariate Gaussian distribution, i.e., Model~\ref{gaussianmodel0}, and Gamma distribution are plotted in panel (a) and (b) in Figure~\ref{treeplot}, respectively. Each dataset is with binary tree topology and each node has one observation. The parameter settings in either panel are the same except the correlation. The detailed procedures to generate data are displayed in Section~\ref{sec:simul}. In each panel, we can visually find that, in either sub-figure, the points of later generations have higher degree of dispersion, and between the two sub-figures in the same generation, the data points with smaller correlation are more widely distributed. Therefore, under the condition that all the other parameters are the same, just as shown in Figure~\ref{treeplot}, smaller correlation parameter in corresponding model, like $\rho$ in Model~\ref{gaussianmodel0}, may require a larger angle included by those two lines. And this finding may be independent of the model.
\newsavebox{\smlmatdistributionb}
\savebox{\smlmatdistributionb}{$N \begin{pmatrix} \begin{pmatrix} 2 \\ 2 \end{pmatrix} &, & \begin{pmatrix} 1.5 & 1.5 \rho^i \\ 1.5 \rho^i & 1.5 \end{pmatrix} \end{pmatrix}$}
\begin{figure}
	\centering
	\includegraphics[scale=0.25]{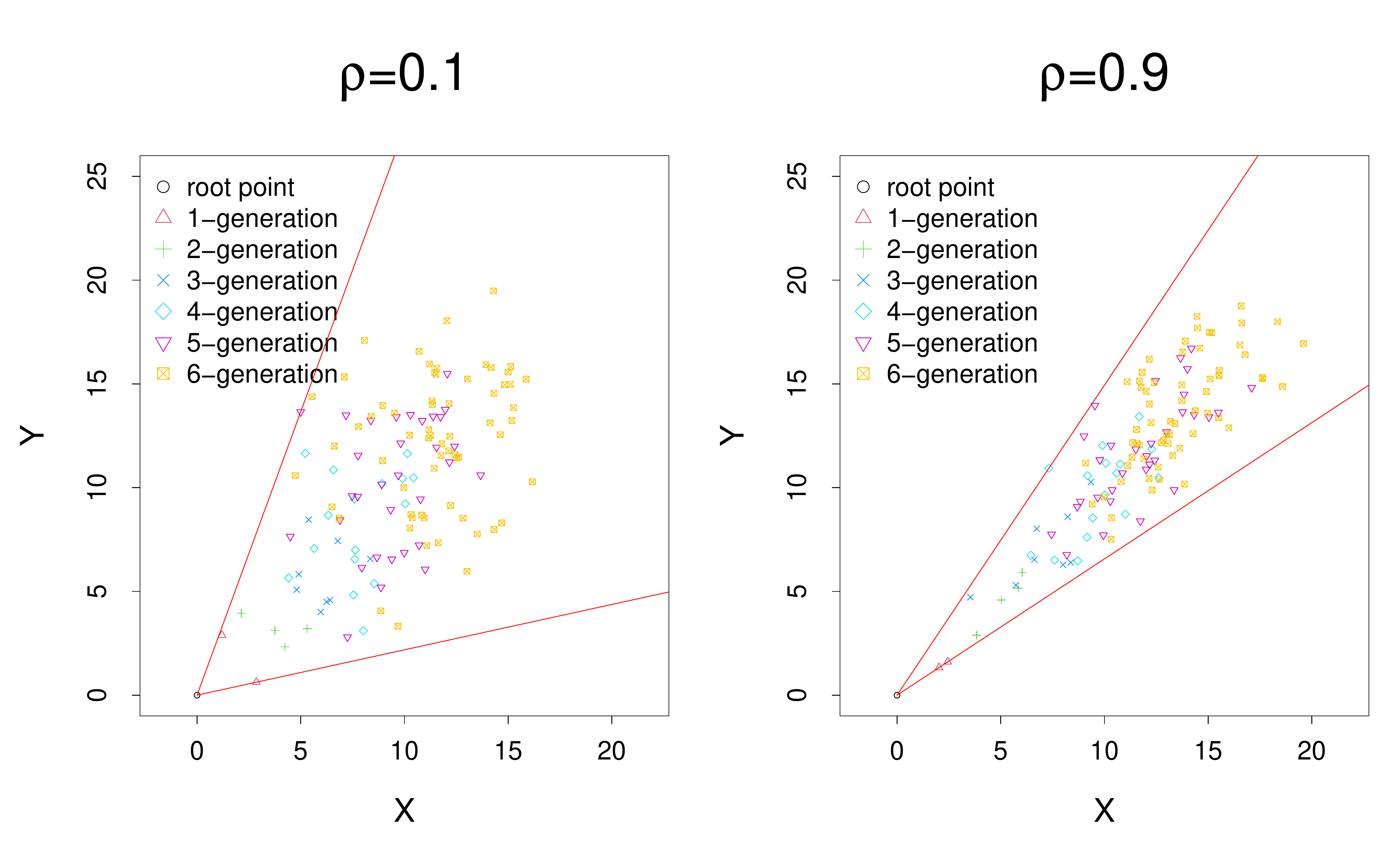}
	
	\centering{(a)}
	
	\includegraphics[scale=0.25]{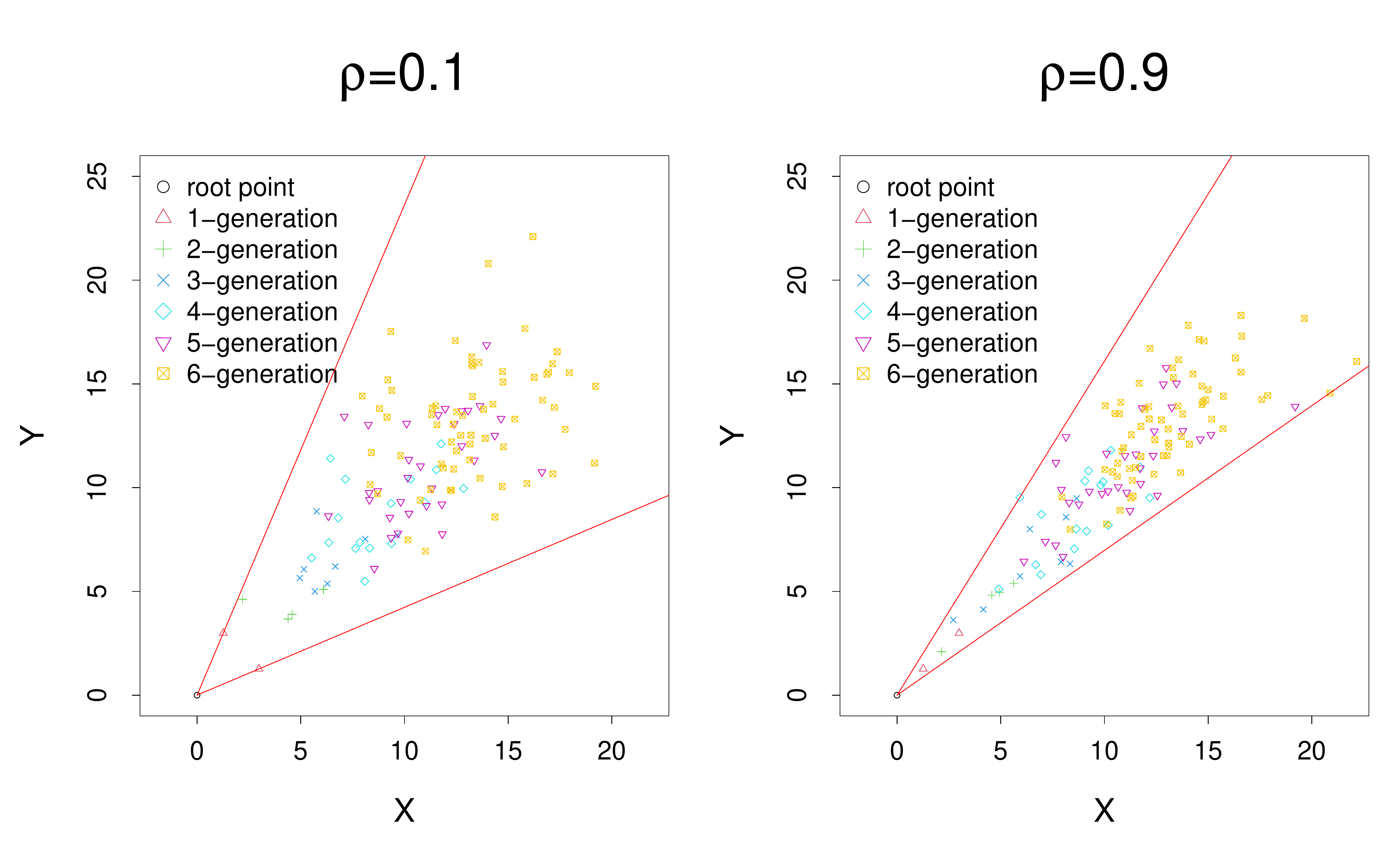}
	
	\centering{(b)}
	\caption{Scatter plot of samples. In panel (a) and (b), the datasets are generated from binary Model~\ref{gaussianmodel0} with the increments following \usebox{\smlmatdistributionb} and Gamma distribution, respectively, where the correlation $\rho$ is labeled at the top. In each sub-figure, two lines which intersect at fixed root point $(0,0)$ and pack all observations with the minimum included angle are plotted.}
	\label{treeplot}
\end{figure}

\begin{figure}
	\vspace{-0.8cm}
	\centering
	\includegraphics[scale=0.17]{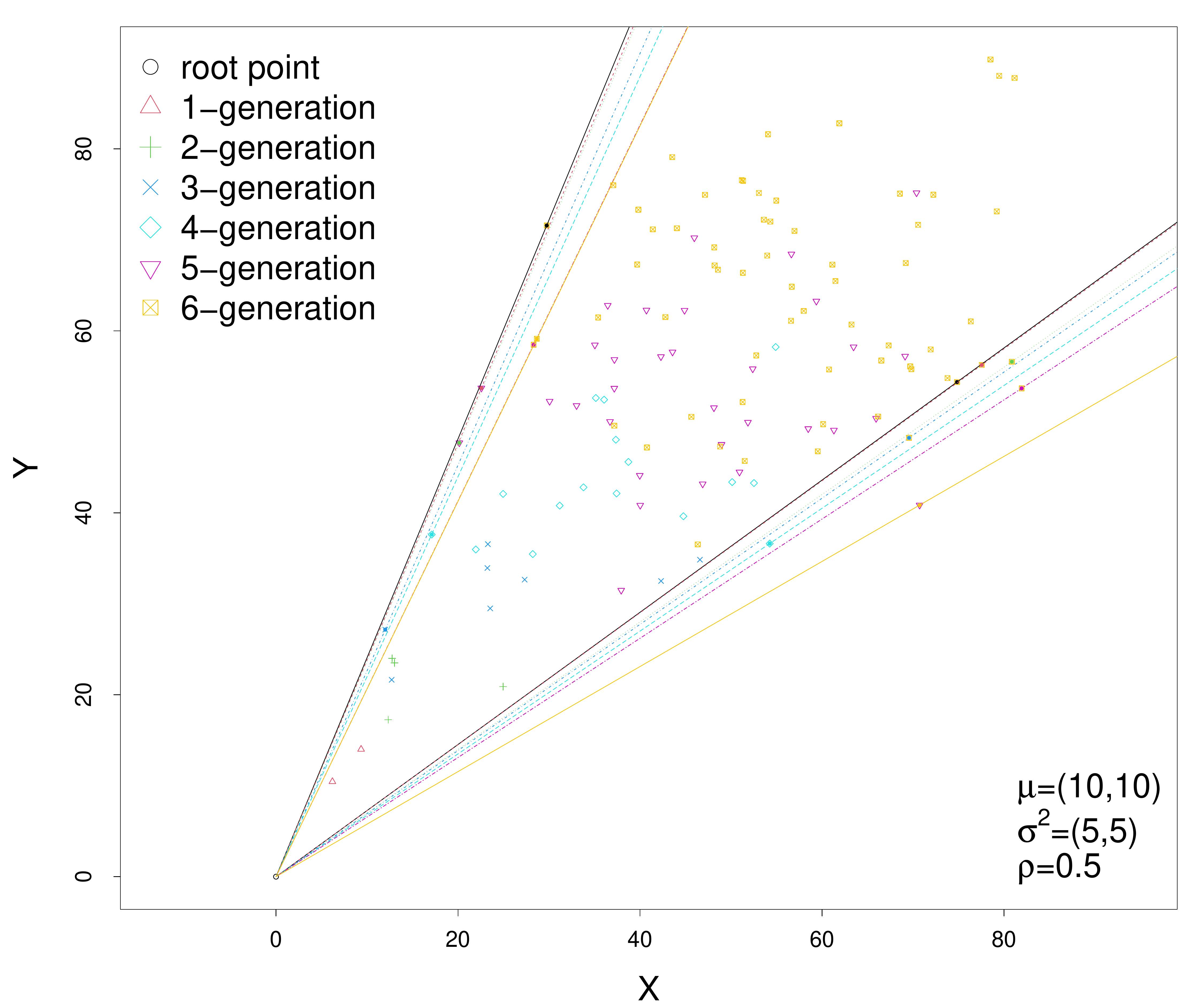}
	\caption{All candidate pairs of lines based on a synthesized tree-shaped dataset. Each angle between a pair of lines is one candidate for $\Delta \hat{\theta}$.}
	\label{allangles}
\end{figure}

Under this condition, this kind of angle, denoted as $\Delta \theta$, can initially be used to quantify the correlation between tree-shaped datasets. Thus, how to obtain the estimation of this angle, denoted as $\Delta \hat{\theta}$, is a very important question and here is an intuitive way. With fixed intersection point, usually the origin point, each of the two lines should go through at least one observation point and the proportion of the observations located between the lines is at least 95\% (including the points on the lines which is called side-point). The included angles in Figure~\ref{treeplot}, which include all points, are very sensitive to outliers, leading to a large variance. Therefore, the scheme that includes 95\% points is adopted. Figure~\ref{allangles} displays all candidate pairs of lines. There are many ways to get $\Delta \hat{\theta}$ from these candidates, such as taking the mode, or the median. Finally, the average value of these candidates is adopted as $\Delta \hat{\theta}$, because of its small variance.

The simulation results in Section~\ref{sec:simul} show that, with the same parameters except the correlation, the proposed $\Delta \theta$ can well measure the relative magnitude of correlation between tree-shaped datasets, even if they follow non-Gaussian or discrete distribution. In the following content, $\Delta \theta$ will be proved to be a suitable and valid geometric statistic, similar to Pearson correlation coefficient, and can be used to display the tree correlation well. A normalization algorithm to get more accurate $\Delta \hat{\theta}$ is also proposed.

\section{Semi-parametric Model}
\label{sec:model}

To generalize Model~\eqref{gaussianmodel0}, the following new model is introduced:
\begin{footnotesize}
	\begin{equation}
		\begin{split}
			& \begin{pmatrix}
				X^{i,j}_{T^{A(\cdot,i,j)}+0} \\ Y^{i,j}_{T^{A(\cdot,i,j)}+0}
			\end{pmatrix}
			=
			\begin{pmatrix}
				X^{A(i-1,i,j)}_{T^{A(\cdot,i,j)}} \\ Y^{A(i-1,i,j)}_{T^{A(\cdot,i,j)}}
			\end{pmatrix} \ ,\quad \text{ for } i =2, 3, \cdots;\\
			& \left(\begin{array}{c|c}
				X^{i,j}_{T^{A(\cdot,i,j)}+t}& \\
				Y^{i,j}_{T^{A(\cdot,i,j)}+t}&
			\end{array} \begin{array}{cc}
				X^{i,j}_{T^{A(\cdot,i,j)}+t-1} = x^{i,j}_{T^{A(\cdot,i,j)}+t-1}& \\
				Y^{i,j}_{T^{A(\cdot,i,j)}+t-1} = y^{i,j}_{T^{A(\cdot,i,j)}+t-1}&
			\end{array} \right)
			\sim N
			\begin{pmatrix}
				\begin{pmatrix}
					\mu_i^x + x^{i,j}_{T^{A(\cdot,i,j)}+t-1} \\ \mu_i^y + y^{i,j}_{T^{A(\cdot,i,j)}+t-1}
				\end{pmatrix}
				&
				,
				&
				\begin{pmatrix}
					\sigma_{1i}^2 & f(i; \rho) \sigma_{1i} \sigma_{2i} \\
					f(i; \rho) \sigma_{1i} \sigma_{2i} & \sigma_{2i}^2
				\end{pmatrix}
			\end{pmatrix}, \\
			& \quad \text{ for } i = 1, 2, \cdots, \ \ \ t = 1, 2, \cdots, T^{i,j},
		\end{split}
		\label{gaussianmodel}
	\end{equation}
\end{footnotesize}
\noindent where $f(i;\rho) > 0$ is a decreasing function of generation $i$ and an increasing function of $\rho$. This model is called semi-parametric Gaussian model (SPGM). Specifically, the first formula represents the dependence between parent-child nodes by treating $X^{i,j}_{T^{A(\cdot,i,j)}+0}, Y^{i,j}_{T^{A(\cdot,i,j)}+0}$ as two auxiliary random variables equal to $X^{A(i-1,i,j)}_{T^{A(\cdot,i,j)}}, Y^{A(i-1,i,j)}_{T^{A(\cdot,i,j)}}$, respectively. In the second formula, random walk is used to model the dependence between neighboring observations within a node. The association between two trees is modeled by the positively correlated increments, with the correlation $f(i; \rho)$ weakening as the tree depth $i$ increasing.

In term of the damping correlation, $f(i; \rho)$, it is motivated by some real problems. For example, in cell lineage tree, individual difference may be larger and larger as the generation increasing, so that correlation between individuals is weakening \citep{kaneko1994cell}. Moreover, without lose of generality, both of $f(i;\rho)$ and $\rho$ are constrained to be positive. In this paper, two examples of $f(i;\rho)$ are adopted in following contents: one is $f(i;\rho)=\rho^i$, which is often encountered in statistical modeling \citep{haynes1974application, nakazato1996temporal, fery2005physical}; the other common one is linear form $f(i;\rho)=(1-\frac{i-1}{I}) \rho$, where $I=max(i)$ \citep{Fiore2020, MITRA2002556, 9404292}. Of course, other patterns can be proposed case by case.

Sometimes, there's only one new observation for each node, such as the lifetime of each cell in a cell lineage tree \citep{murray2012multidimensional}. This situation can be treated as a special case of SPGM with $T^{i,j}=1$ (hence omit the subscript $t$) and $\mu^x_{i}=\mu^x, \mu^y_{i}=\mu^y, \sigma_{1i}=\sigma_1, \sigma_{2i}=\sigma_2$ for all $i,j$, which is called the degenerated SPGM (DSPGM). Let $X^{i,j}$ and $Y^{i,j}$ be the only new observation in node $(i,j)$ of tree $\boldsymbol{X}$ and $\boldsymbol{Y}$, respectively, then, the DSPGM is simplified as follows:
\begin{small}
	\begin{equation}
		\begin{split}
			& \left(\begin{array}{c|c}
				X^{i,j}& \\
				Y^{i,j}&
			\end{array} \begin{array}{cc}
			X^{A(i-1,i,j)} = x^{A(i-1,i,j)}& \\
			Y^{A(i-1,i,j)} = y^{A(i-1,i,j)}&
		\end{array} \right)
			\sim N
			\begin{pmatrix}
				\begin{pmatrix}
					\mu^x + x^{A(i-1,i,j)} \\ \mu^y + y^{A(i-1,i,j)}
				\end{pmatrix}
				,
				\begin{pmatrix}
					\sigma_1^2 & f(i; \rho) \sigma_1 \sigma_2 \\ f(i; \rho) \sigma_1 \sigma_2 & \sigma_2^2
				\end{pmatrix}
			\end{pmatrix},\\
		\end{split}
		\label{degeneratedmodel}
	\end{equation}
\end{small}
where $i = 2,3,\cdots$.

\section{Geometric Interpretation}
\label{sec:graph}

According to the description in Section~\ref{sec:pre}, $\Delta \theta$ may reflect the relative magnitude of $f(i;\rho)$, which is the main concern in the paper. Before giving the detailed proof, some background knowledge are provided at first.

\subsection{Quantile Ellipse}
Since the support set of bivariate Gaussian distribution locates at $\mathbb{R}^2$, it is impossible to find two lines to pack all samples with probability 1. Therefore, the quantile ellipse is proposed to find an elliptical boundary centered at mean point of bivariate Gaussian distribution with fixed quantile.

Let's first introduce the definition of quantile ellipse \citep{Husson2005Confidence} of the following bivariate Gaussian distribution.
\begin{small}
	\begin{equation}
		\begin{split}
			\begin{pmatrix}
				X \\ Y
			\end{pmatrix}
			\sim N
			\begin{pmatrix}
				\begin{pmatrix}
					\mu_1 \\ \mu_2
				\end{pmatrix}
				&
				\begin{pmatrix}
					\sigma_{1}^2 & \rho\sigma_{1}\sigma_{2} \\ \rho\sigma_{1}\sigma_{2} & \sigma_{2}^2
				\end{pmatrix} 
			\end{pmatrix}.
		\end{split}
		\label{twonorm}
	\end{equation}
\end{small}

\begin{figure}
	\vspace{-0.8cm}
	\centering
	\includegraphics[scale=0.57]{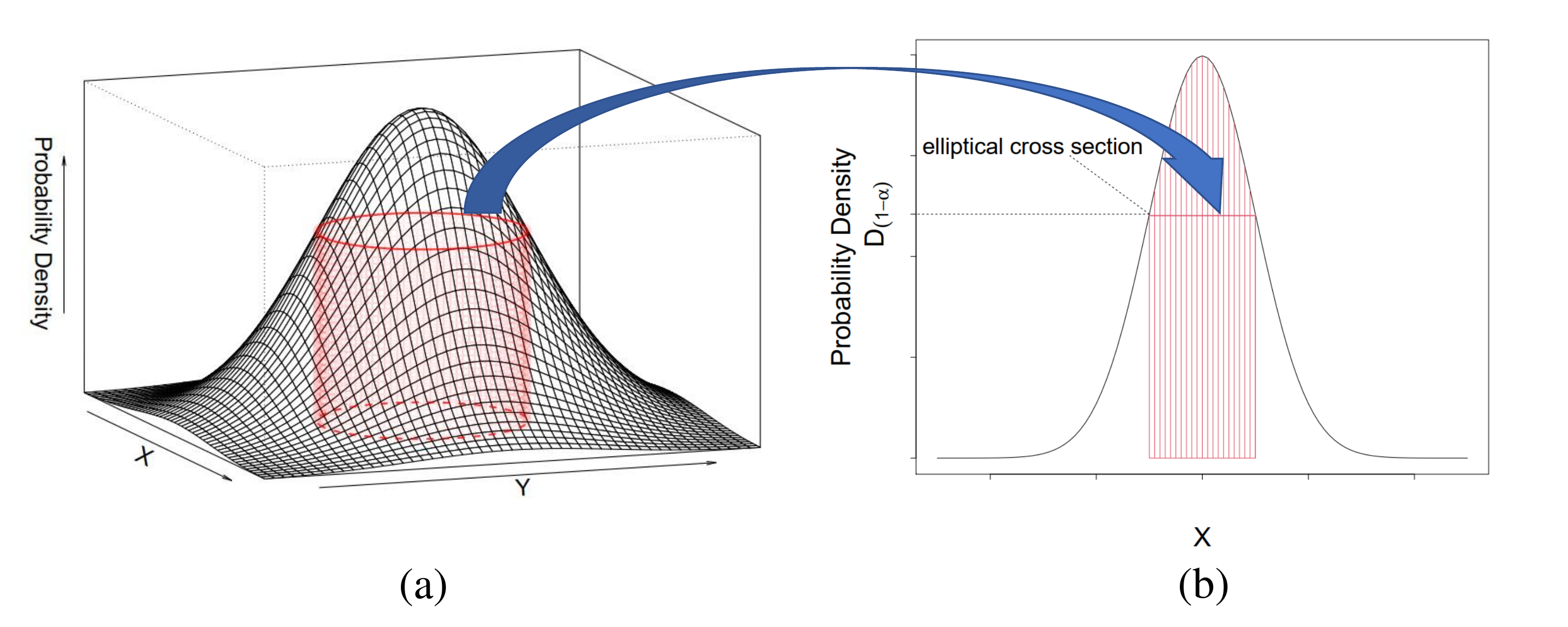}
	\caption{Schematic high density area of a bivariate Gaussian distribution. The left panel is a 3D plot of a bivariate Gaussian, and the right panel is its corresponding schematic high density section.}
	\label{ellipse95}
\end{figure}
The contour of the density function of distribution \eqref{twonorm} is an ellipse centered at $\begin{pmatrix} \mu_1 \\ \mu_2 \end{pmatrix}$. And the corresponding cylindrical surface, i.e.,
\begin{equation*}
	\begin{cases}
		\frac{1}{2\pi\sigma_1\sigma_2\sqrt{1-\rho^2}}e^{-\frac{1}{2(1-\rho^2)} \{ (\frac{x-\mu_1}{\sigma_1})^2 - 2\rho(\frac{x-\mu_1}{\sigma_1})( \frac{y-\mu_2}{\sigma_2}) + (\frac{y-\mu_2}{\sigma_2})^2 \}} = D_{(1-\alpha)}\\
		z > 0\\
	\end{cases},
\end{equation*}
divides the space between probability density curved surface and $x-y$ plane into two parts: the inside part which is inside the surface and the outside part that is outside the surface. As Figure~\ref{ellipse95}(b) shown, the striated part and non-striated part under the probability density surface represent the inside part and outside part, respectively. In this figure, the height of the elliptical cross section of probability density surface and the cylindrical surface, i.e. $D_{(1-\alpha)}$, is chosen to make the volume of the striated part equal to $(1-\alpha)$, and this ellipse is defined as the $(1-\alpha)$ quantile ellipse of the Gaussian distribution. Through calculation provided in Section~S1 of the supplementary material, we get:
\[ D_{(1-\alpha)} = \frac{\alpha}{2\pi\sigma_1\sigma_2\sqrt{1-\rho^2}}. \]
Let the projection of $(1-\alpha)$ quantile ellipse to $x-y$ plane be:
\begin{equation}
\begin{cases}
(\frac{x-\mu_1}{\sigma_1})^2 - 2\rho(\frac{x-\mu_1}{\sigma_1})(\frac{y-\mu_2}{\sigma_2}) + (\frac{y-\mu_2}{\sigma_2})^2 = c_{(1-\alpha)}^2\\
z = 0\\
\end{cases}.
\label{quantileellipse}
\end{equation}
Compared with the density function, we get:
\[ D_{(1-\alpha)}=\frac{1}{2\pi\sigma_1\sigma_2\sqrt{1-\rho^2}}e^{-\frac{1}{2(1-\rho^2)} \cdot c_{(1-\alpha)}^2} \ \ , \ \ c_{(1-\alpha)}^2 = -2(1-\rho^2)\ln{\alpha}. \]
This is consistent with the property: If $\boldsymbol{X} \sim \boldsymbol{N}_2(\boldsymbol{\mu},\boldsymbol{\Sigma})$, then the variable $U = (\boldsymbol{X} - \boldsymbol{\mu})^T \boldsymbol{\Sigma}^{-1} (\boldsymbol{X} - \boldsymbol{\mu})$ follows a $\chi^2_2$ distribution \citep{book:805419}.

\begin{figure}
	\vspace{-0.8cm}
	\centering
	\includegraphics[scale=0.28]{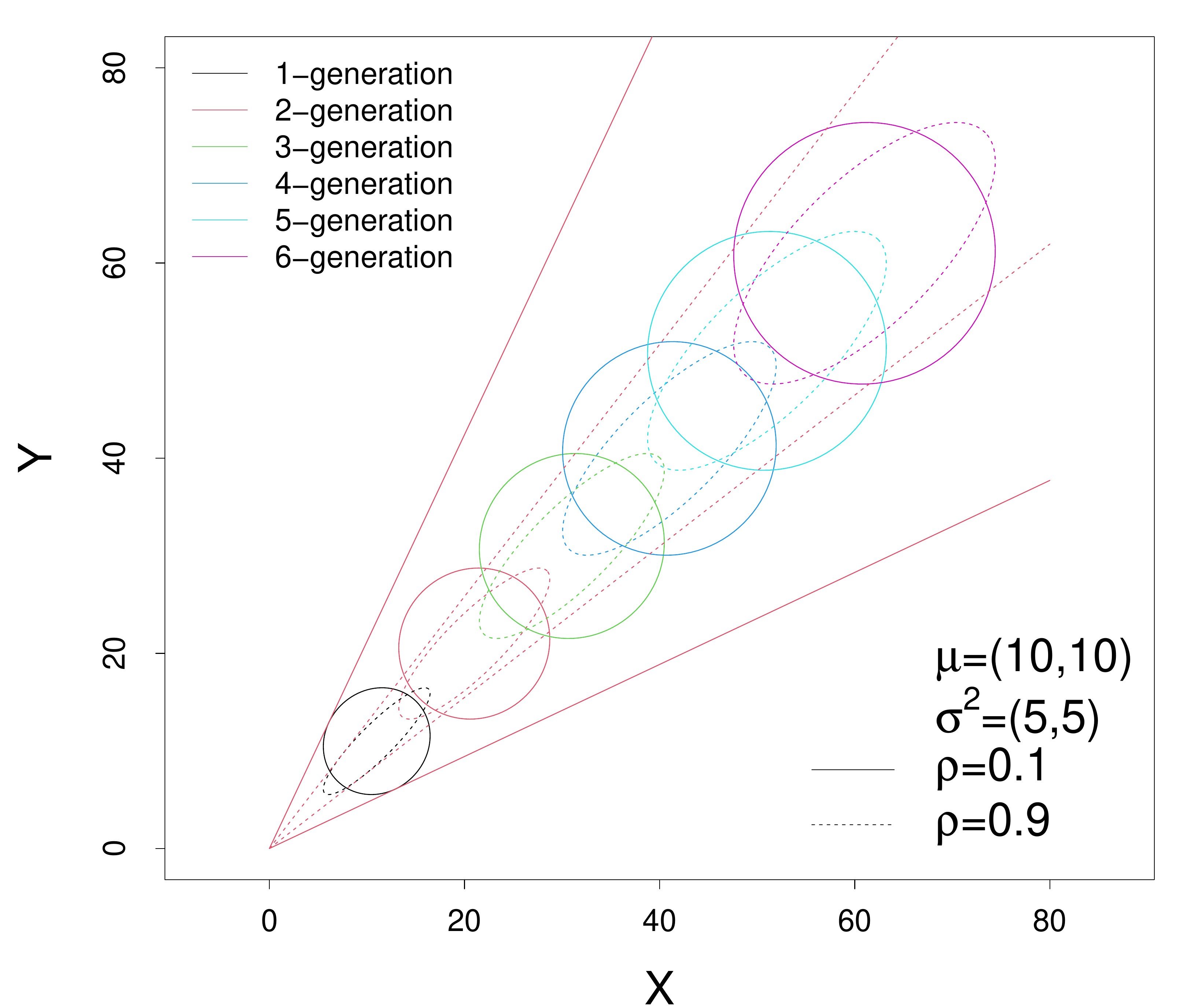}
	\caption{Quantile ellipses of tree-shaped dataset with $f(i;\rho) = \rho^i$. The only difference between distributions corresponding to dotted and solid ellipses is $\rho$.}
	\label{exampleellipse}
\end{figure}

For convenience, $\alpha$ is set to $0.05$ in the following contents. An example of 95\% quantile ellipses based on Model~\eqref{degeneratedmodel} is plotted in Figure~\ref{exampleellipse}. There are two series of ellipses corresponding to two pairs of trees with different $\rho$, and either series has 6 ellipses for 6 generations. The corresponding distribution of each ellipse is the marginal distribution of the corresponding node $(X^{i,j}, Y^{i,j})$,
\begin{small}
	\begin{equation*}
		\begin{split}
			\begin{pmatrix}
				X^{i,j} \\ Y^{i,j}
			\end{pmatrix}
			\sim N
			\begin{pmatrix}
				\begin{pmatrix}
					i \cdot \mu^x \\ i \cdot \mu^y
				\end{pmatrix}
				&
				\begin{pmatrix}
					i \cdot \sigma_{1}^2 & \sum_{r=1}^{i} \rho^r \sigma_{1}\sigma_{2} \\ \sum_{r=1}^{i} \rho^r \sigma_{1}\sigma_{2} & i \cdot \sigma_{2}^2
				\end{pmatrix} 
			\end{pmatrix}.
		\end{split}
	\end{equation*}
\end{small}
Since the marginal distributions of nodes of the same generation are the same, only one ellipse is needed for each generation. There are two pairs of lines, noted as tangent lines, intersecting at the root point and packing all quantile ellipses with smallest angles for either series. Let $\Delta \theta$ be the included angle between two tangent lines of a $(1-\alpha)\%$ quantile ellipse intersecting at a fixed point $(x_0, y_0)$ outside the ellipse. We can find that, for either series, $\Delta \theta$ of the first generation is the target one, and also, $\Delta \theta$ with $\rho = 0.1$ is obviously larger than that with $\rho = 0.9$. Next we are going to prove them theoretically.

\subsection{The Relationship Between $\Delta \theta$ and $f(i; \rho)$}
The two tangent lines of $\Delta \theta$ are called quantile lines and the slopes of two quantile lines are denoted as $k_1, k_2$. Then, following theorems are obtained.
\begin{thm}
	For bivariate Gaussian distribution \eqref{twonorm}, if $(x_0, y_0) \in \{ x_0 < \mu_1-\varepsilon_1, y_0 < \mu_2-\varepsilon_2 \} \cup \{ x_0 > \mu_1+\varepsilon_1, y_0 > \mu_2+\varepsilon_2 \}$, where $\varepsilon_i = \sigma_i \sqrt{-2\ln{\alpha}}, i = 1, 2$ ,then both $k_1, k_2$ are positive.
	\label{thm0}
\end{thm}
The support area of $(x_0, y_0)$ is shown in Figure~S.2 of the supplementary material where the two sub-regions are centrally symmetric about the center of the quantile ellipse. Then $(x_0, y_0)$ is assumed to locate in the region $\{ x_0 < \mu_1-\varepsilon_1, y_0 < \mu_2-\varepsilon_2 \}$ in the following content. We can find that the positive $k_1, k_2$ makes $\Delta \theta$ acute, and moreover, this area is independent of $\rho$, which is proved in Section~S2 of the supplementary material.

Based on the above condition, the relationship between $\Delta \theta$ and $\rho$ is concluded in the following theorem.
\begin{thm}
	For bivariate Gaussian distribution \eqref{twonorm}, if $k_1, k_2$ are positive, then $\Delta \theta$ is a decreasing function of $\rho$.
	\label{thm1}
\end{thm}
Through the proof provided in Section~S3 of the supplementary material, the relationship between $\Delta \theta$ and $\rho$ is given by:
\begin{equation}
\lambda \sigma_1\sigma_2\rho - u_1u_2 = - \sqrt{\lambda^2 \sigma_1^2\sigma_2^2 + u_1^2u_2^2 - \lambda \sigma_1^2u_2^2 - \lambda \sigma_2^2u_1^2 + \frac{\left[ \lambda(\sigma_1^2 + \sigma_2^2) -(u_1^2 + u_2^2) \right]^2}{4} \cdot \tan^2(\Delta\theta)},
\label{anglerho}
\end{equation}
where $u_1 = \mu_1 - x_0, u_2 = \mu_2 - y_0, \lambda = -2 \ln{\alpha}$.

According to Theorem~\ref{thm1}, if all other parameters are given, $\Delta \theta$ is a monotonic decreasing function of $\rho$. Thus, $\Delta \theta$ can be used to demonstrate the geometric interpretation of the correlation $\rho$ in a bivariate Gaussian distribution. However, for tree-shaped dataset, the correlation of $(X^{i,j},Y^{i,j})$ changes with the generation $i$. Let $\Delta \theta_i$ and $k_{1,i}, k_{2,i}$ be the angle between the quantile lines and the two quantile slopes of the $i$-th generation, respectively. The following theorem is proposed to demonstrate the relationship between $\Delta \theta_i$ and $i$, and the proof is provided in Section~S4 of the supplementary material.
\begin{thm}
	Within DSPGM \eqref{degeneratedmodel}, if $\mu^x > 0, \mu^y > 0, k_{1,i} > 0, k_{2,i} > 0$, $\frac{\mu^2}{\lambda\sigma^2} > \max \{ \frac{f(1;\rho)}{\cos 2\gamma}, 1 \}$, where $\cos \gamma = \frac{\sqrt{2}}{2} \frac{\mu^x + \mu^y}{\sqrt{(\mu^x)^2 + (\mu^y)^2}}$, $\sin \gamma = \frac{\sqrt{2}}{2} \frac{\mu^x - \mu^y}{\sqrt{(\mu^x)^2 + (\mu^y)^2}}$, $\sigma^2 = \max (\sigma_1^2, \sigma_2^2) \cdot \cos 2\gamma$, $\mu = \frac{\sigma}{\sigma_1}\mu^x\cos \gamma - \frac{\sigma}{\sigma_2}\mu^y\sin \gamma$, then $\Delta \theta_i$ is a decreasing function of generation $i$.
	\label{thm2}
\end{thm}

Based on Theorem~\ref{thm1} and Theorem~\ref{thm2}, we can conclude that, under some conditions, the quantile lines of the first generation definitely pack all the $(1-\alpha)$\% quantile ellipses of all the other generations. That means observations of later generations, which constitute majority of samples due to the tree structure, may be ignored when searching for $\Delta \theta$. Besides, the quantile ellipses in Figure~\ref{exampleellipse} are corresponding to the unknown marginal distributions of $(X^{i,j},Y^{i,j})$, while in practice only conditional distributions of $(X^{i,j},Y^{i,j} | X^{A(i-1,i,j)},Y^{A(i-1,i,j)})$ are sampled, and even the conditional distributions is changing with $(i,j)$. Meanwhile, other parameters, except for $\rho$, may also influence the magnitude of $\Delta \theta$. In term of these difficulties, a normalization algorithm, called $TD \Delta \theta$, is proposed.

\subsection{$TD \Delta \theta$ Algorithm}

Since the increments, $\{ X^{i,j}_{T^{A(\cdot,i,j)}+t+1} - X^{i,j}_{T^{A(\cdot,i,j)}+t}, Y^{i,j}_{T^{A(\cdot,i,j)}+t+1} - Y^{i,j}_{T^{A(\cdot,i,j)}+t} \}$ in SPGM and  and $\{ X^{i,j} - X^{A(j-1,j,i)}, Y^{i,j} - Y^{A(j-1,j,i)} \}$ in DSPGM, are independent and identically distributed within the same generation, they can be adopted directly, rather than the original data $( \boldsymbol{X}, \boldsymbol{Y} )$, to calculate the sample $\Delta \theta$, denoted as $\Delta \hat{\theta}$. Next, a normalization procedure on the increments is proposed to solve the dilemma of unknown marginal distribution and eliminate the impact of generation $i$ and other parameters to the most extent.

In the following steps, let $(\Delta \boldsymbol{X}_e^i, \Delta \boldsymbol{Y}_e^i)$ be the increments of the $i$-th generation after Step $e$. It's worth mentioning that all transformations in the normalization procedure are linear, and hence do not change the correlation. The detailed normalization steps, named Tree-shaped Datasets $\Delta \theta$ ($TD \Delta \theta$), is introduced as follows:
\begin{itemize}
	\item[Step 1:] Let original data minus the root values to make the root be origin point and obtain increments $\{ (\Delta \boldsymbol{X}_1^i, \Delta \boldsymbol{Y}_1^i); i =1,2,\cdots \}$;
	\item[Step 2:] Normalize $\{ (\Delta \boldsymbol{X}_1^i, \Delta \boldsymbol{Y}_1^i); i =1,2,\cdots \}$ to have the same variance $\sigma^2$ and the given mean $\mu_i^*$ (this step is skipped if other parameters of two pairs of trees are known to be the same except $\rho$):
		\begin{itemize}
			\item[Step 2.1:] Use $\{ (\Delta \boldsymbol{X}_1^i, \Delta \boldsymbol{Y}_1^i); i =1,2,\cdots \}$ to obtain estimators $\hat{\mu}_i^x, \hat{\mu}_i^y, \hat{\sigma}_{1i}^2$, $\hat{\sigma}_{2i}^2$, $i =1,2,\cdots$, by Maximum Likelihood Estimate (MLE);
			\item[Step 2.2:] Let $(\Delta \boldsymbol{X}_{2.1}^i, \Delta \boldsymbol{Y}_{2.1}^i)$, $i =1,2,\cdots$, multiply $( \frac{\sigma}{\hat{\sigma}_{1i}}, \frac{\sigma}{\hat{\sigma}_{2i}} )$, where $\sigma$ is predetermined;
			\item[Step 2.3:] Let $(\Delta \boldsymbol{X}_{2.2}^i, \Delta \boldsymbol{Y}_{2.2}^i)$, $i =1,2,\cdots$, minus $( \hat{\mu}_i^x, \hat{\mu}_i^y )$ and then plus $\mu_i^*$, where $\mu_i^*$ is predetermined;
		\end{itemize}
	\item[Step 3:] Fix the intersect point $(x_0, y_0)$ to the origin and use $\{ (\Delta \boldsymbol{X}_2^i, \Delta \boldsymbol{Y}_2^i); i =1,2,\cdots \}$ to get $\Delta \hat{\theta}$.
\end{itemize}

\subsubsection{Determination of $\sigma$ and $\mu_i^*$}
After Step 2, the original increments $(\Delta \boldsymbol{X}, \Delta \boldsymbol{Y})$ can be transformed into $(\Delta \boldsymbol{X}_*, \Delta \boldsymbol{Y}_*)$ following:
\begin{small}
	\begin{equation}
		\begin{split}
			\begin{pmatrix}
				X^{i,j}_{T^{A(\cdot,i,j)}+t+1 *} - X^{i,j}_{T^{A(\cdot,i,j)}+t *} \\ Y^{i,j}_{T^{A(\cdot,i,j)}+t+1 *} - Y^{i,j}_{T^{A(\cdot,i,j)}+t *}
			\end{pmatrix}
			\sim N &
			\begin{pmatrix}
				\begin{pmatrix}
					\mu_i^* \\ \mu_i^*
				\end{pmatrix}
				&
				,
				&
				\begin{pmatrix}
					\sigma^2 & f(i;\rho) \sigma^2 \\ f(i;\rho) \sigma^2 & \sigma^2
				\end{pmatrix}
			\end{pmatrix} \ \ \ \text{for SPGM}, \\
			\begin{pmatrix}
				X^{i,j}_* - X^{A(j-1,j,i)}_* \\ Y^{i,j}_* - Y^{A(j-1,j,i)}_*
			\end{pmatrix}
			\sim N &
			\begin{pmatrix}
				\begin{pmatrix}
					\mu_i^* \\ \mu_i^*
				\end{pmatrix}
				&
				,
				&
				\begin{pmatrix}
					\sigma^2 & f(i;\rho) \sigma^2 \\ f(i;\rho) \sigma^2 & \sigma^2
				\end{pmatrix}
			\end{pmatrix} \ \ \ \text{for DSPGM}. \\
		\end{split}
		\label{normalizemodel}
	\end{equation}
\end{small}

In fact, the location of these quantile ellipses, which is only related to the $\mu_i^*$, is easier to manipulate than the shape of these ellipses. Therefore, all normalized increments of different generations share the same given variance $\sigma^2$, and the expectation of each generation $\mu_i^*$ is adjusted generation by generation. In detail, these $\mu_i^*$ are set by the following process. Let $\Delta \theta_i$ be the included angle between the quantile lines of generation $i$.

For the $1$-st generation, via Equation~\eqref{anglerho} and Model~\eqref{degeneratedmodel}, $\Delta \theta_1$ follows:
\[ \sqrt{1+\tan^2 (\Delta \theta_1)} = \frac{\mu_1^{*2} - \lambda\sigma^2 \rho}{\mu_1^{*2} - \lambda\sigma^2} = 1+\frac{1-\rho}{\frac{\mu_1^{*2}}{\lambda\sigma^2} - 1}. \]
The absolute value of the derivative of $\sqrt{1+\tan^2 (\Delta \theta_1)}$ with respect to $\rho$ is $| \frac{d \sqrt{1+\tan^2 (\Delta \theta_1)}}{d \rho} | = \frac{1}{\frac{\mu_1^{*2}}{\lambda\sigma^2} - 1}$. The bigger this absolute value is, the more sensitive $\sqrt{1+\tan^2 (\Delta \theta_1)}$ is to the change of $\rho$. Thus, let $\mu_1^{*2} = \lambda\sigma^2 + \tau$, where $\tau$ is a small positive number, so that the absolute value of the derivative could be large enough.

\begin{figure}[htb]
	\centering
	\includegraphics[scale=0.2]{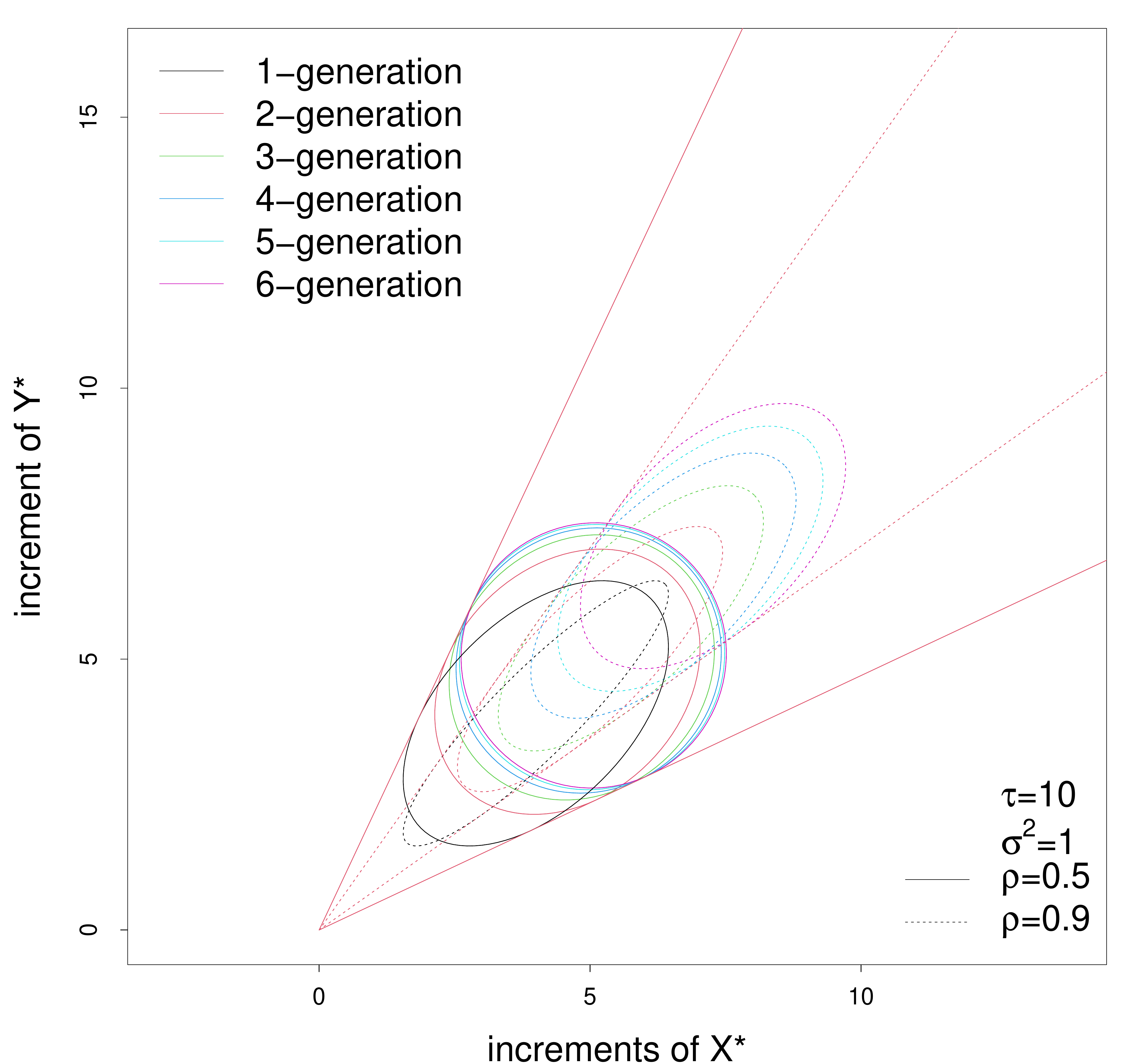}
	\caption{Quantile ellipses for the tree-shaped dataset after normalization with $f(i;\rho)=\rho^i$. Either series of quantile lines are common to their corresponding series of ellipses.}
	\label{normalizedellipse}
\end{figure}

For the $i$-th generation, $\Delta \theta_i$ should satisfy:
\[ \sqrt{1+\tan^2 (\Delta \theta_i)} = 1+\frac{1-f(i;\rho)}{\frac{\mu_i^{*2}}{\lambda\sigma^2} - 1}. \]
It would be ideal if $\Delta \theta_i$ are the same for different $i$, and hence all the quantile ellipses of the increments have common tangent lines, as shown in Figure~\ref{normalizedellipse}. Therefore, all increments are important in calculating $\Delta \hat{\theta}$ despite of the damping correlation system. If all $\Delta \theta_i$ are equal, the following equality should be held:
\[ \frac{1-f(i;\rho)}{\frac{\mu_i^{*2}}{\lambda\sigma^2} - 1} = \frac{1-\rho}{\frac{\mu_1^{*2}}{\lambda\sigma^2} - 1} \Rightarrow \mu_i^{*2} = \frac{1-f(i;\rho)}{1-\rho} \tau + \lambda\sigma^2. \]
However, the pattern $f$ is usually unknown in practice, so an approximation of $\frac{1-f(i;\rho)}{1-\rho}$ should be estimated, noted as $\epsilon_i$. We get $\mu_i^{*2} = \epsilon_i \tau + \lambda \sigma^2$. Our suggestion is to set $\epsilon_i = \sum_{j=1}^{i}\frac{1}{j}$ which performs well in simulation study.

\subsubsection{Calculation of $\Delta \hat{\theta}$}

After normalization, the tree-shaped dataset satisfies Theorem~\ref{thm0}, since the distribution of increments follows Model~\eqref{normalizemodel} and we can get:
\[ \mu_1 - \varepsilon_1 = \mu_i^* - \sigma\sqrt{\lambda} = \sqrt{\epsilon_i\tau + \lambda \sigma^2} - \sigma\sqrt{\lambda} > 0 \]
Similarly, we can get $\mu_2 - \varepsilon_2 > 0$.

\begin{figure}[htb]
	\vspace{-0.8cm}
	\centering
	\includegraphics[scale=0.4]{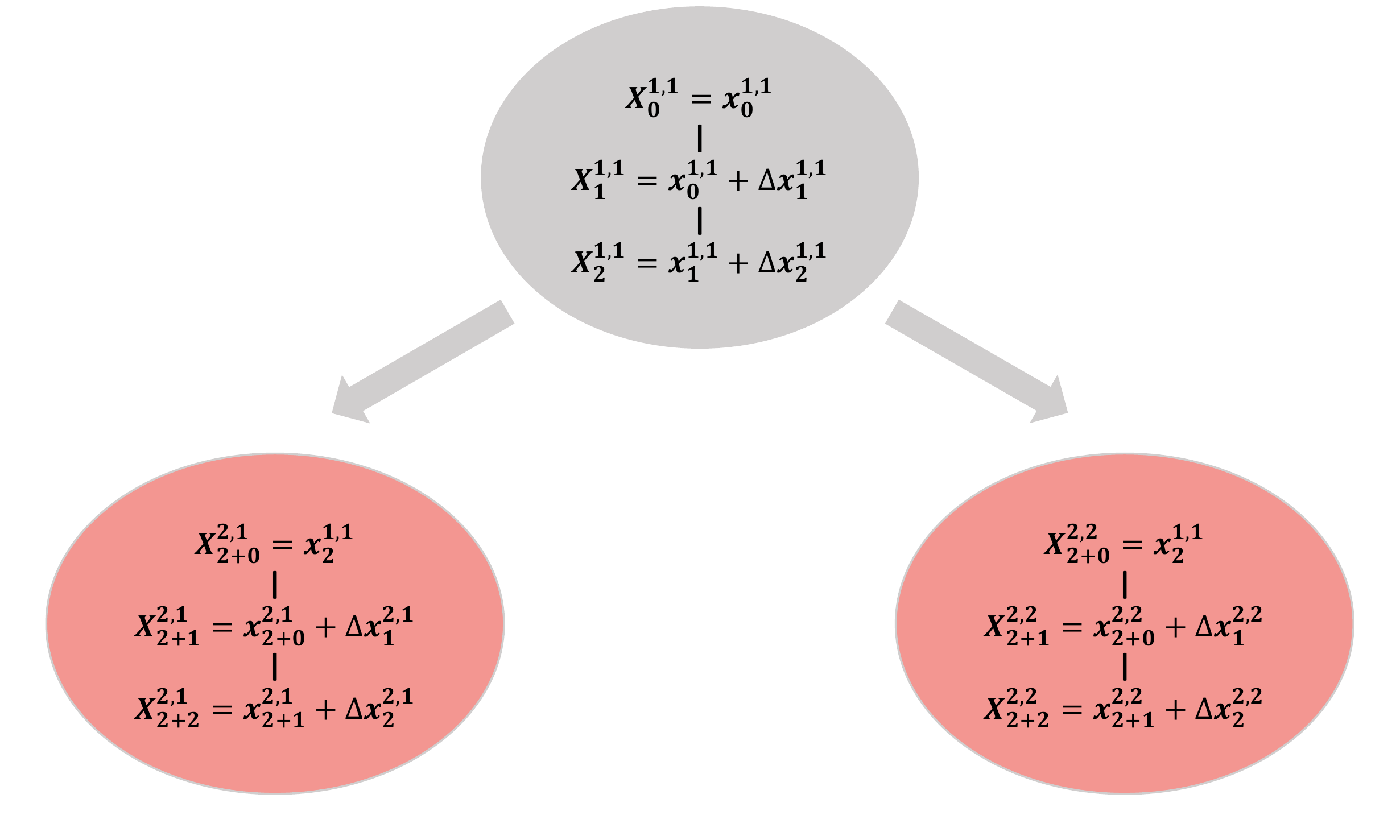}
	
	\begin{Huge}\centering{$\Downarrow$}\end{Huge}
	
	\includegraphics[scale=0.5]{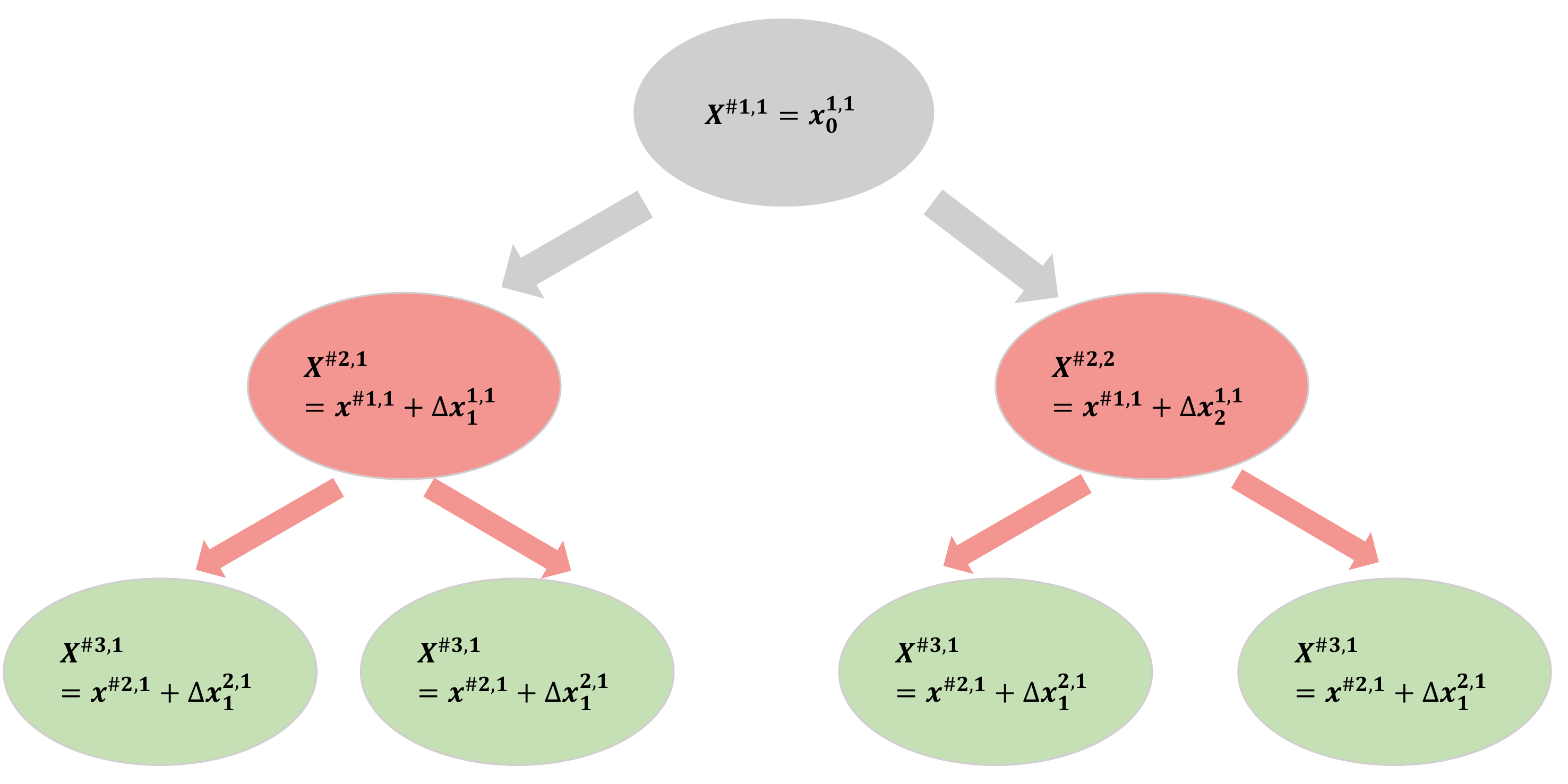}
	\caption{An example of the transformation from SPGM $\boldsymbol{X}$ into DSPGM $\boldsymbol{X}^{\#}$.}
	\label{GMtoDM}
\end{figure}
Under this condition, of the calculation of $\Delta \hat{\theta}$ for a real dataset, recalling Figure~\ref{treeplot}, the method proposed in Section~\ref{sec:pre} can be adopted. It is worth mentioned that the data used to calculate $\Delta \hat{\theta}$ is increments, there are two advantages for this: 1. all increments are independent, so that the impact of conditional distributions can be eliminated; 2. SPGM can actually be transformed into DSPGM without changing the correlation. In other words, node $(i,j)$ in SPGM which contains a time series of length $T^{i,j}$ can be replaced by $T^{i,j}$ nodes in DSPGM where each node inherits one increment from the corresponding time series in SPGM. An example of tree $\boldsymbol{X}$ from SPGM transformed into $\boldsymbol{X}^{\#}$ from DSPGM is given in Figure~\ref{GMtoDM}.

\section{Simulation Results}
\label{sec:simul}
In this section, extensive simulations are conducted to assess the $TD \Delta \theta$ algorithm. Besides Gaussian distribution, non-Gaussian distributions and discrete distributions are also generated by Copula function \citep{2000David} to verify our method on various types of tree-shaped datasets. The following steps are applied to assess the proposed method:
\begin{itemize}
	\setlength{\itemsep}{0pt}
	\setlength{\parsep}{0pt}
	\setlength{\parskip}{0pt}
	\item[Step 1:] Given $\rho$ and $\eta$, let $f_1(i;\rho) = \rho^i$, $f_2(i;\rho+\eta) = (\rho + \eta)^i$, other parameters are drawn randomly;
	\item[Step 2:] Synthesize two pairs of tree-shaped datasets $(\boldsymbol{X_1}, \boldsymbol{Y_1})$ and $(\boldsymbol{X_2}, \boldsymbol{Y_2})$ based on specific model with parameters generated in Step 1;
	\item[Step 3:] Calculate two included angles $\Delta \theta_1$, $\Delta \theta_2$ by $TD \Delta \theta$ algorithm;
	\item[Step 4:] Repeat Step 1,2,3 for 1000 times and calculate the proportion that $\Delta \theta_1 > \Delta \theta_2$;
	\item[Step 5:] Repeat Step 4 for 100 times to get the mean and standard deviation of the proportion calculated in Step 4.
\end{itemize}
In Step 1, we select $\rho$ from $\{ 0.1, 0.2, \cdots, 0.9 \}$, then $\eta$ from $\{ 0.05, 0.15, \cdots, 0.85 \}$ with constraint $\rho + \eta < 1$. In Step 2, all datasets are binary trees with 7 generations and the specific steps for generating particular types of data are detailed in the corresponding parts below. Simulations are conducted under two different settings: the other parameters are kept the same for two pairs of trees; or the other parameters are different for two pairs of trees. In either setting, both experiments with and without $TD \Delta \theta$ normalization are performed. Here, we set $\tau = 0.1, \sigma^2 = 1, \epsilon_i = \sum_{j=1}^{i}\frac{1}{j}$.

Moreover, simulation results with $f_1(i;\rho)=(1-\frac{i-1}{I}) \rho$, $f_2(i;\rho+\eta)=(1-\frac{i-1}{I}) (\rho+\eta)$, where $I=max(i)$, are displayed in Section~S5 of the supplementary material.

\subsection{Gaussian Data}
Large scale simulations are performed on tree-shaped dataset following Gaussian distribution. The detailed process to generate datasets is as follows:
\begin{itemize}
	\item[ ]
		\begin{itemize}
			\setlength{\itemsep}{0pt}
			\setlength{\parsep}{0pt}
			\setlength{\parskip}{0pt}
			\item[Step 2.1:] Generate the increments $(\Delta \boldsymbol{X}, \Delta \boldsymbol{Y})$ based on Model~\eqref{degeneratedmodel};
			\item[Step 2.2:] Given the root point $(X^0, Y^0)$, according to $(\Delta \boldsymbol{X}, \Delta \boldsymbol{Y})$, construct the first pair of tree-shaped datasets $(\boldsymbol{X}_1, \boldsymbol{Y}_1)$;
			\item[Step 2.3:] Do the same to generate the second pair of tree-shaped datasets $(\boldsymbol{X}_2, \boldsymbol{Y}_2)$.
		\end{itemize}
\end{itemize}

In Table~\ref{graph_simul_same_diff}, the mean proportions and standard deviations in brackets are displayed, for each $\rho$, there are two rows corresponding to the results without and with normalization, respectively. We can find that the larger $\eta$, or the larger $\rho$, the higher the accuracy. This result indicates that our proposed $TD \Delta \theta$ algorithm can detect the relative degree of the correlation between tree-shaped datasets to a certain extent.
\begin{table}[!h]
	\vspace{-0.8cm}
	\centering
	\caption{Simulation results based on the Gaussian distribution.}
	\resizebox{5.7in}{1.6in}{
	\resizebox{\linewidth}{!}{
		\begin{tabular}{c|c|c|c|c|c|c|c|c|c|c|c|c}
			\multicolumn{10}{c}{\Large Same parameters} & \multicolumn{3}{c}{} \\
			\hline
			\bottomrule[2pt]
			\backslashbox{$\rho$}{$\eta$} & 0.05 & 0.15 & 0.25 & 0.35 & 0.45 & 0.55 & 0.65 & 0.75 & 0.85 & & \diagbox[dir=SW]{~~~~~~}{~~~~~~} & \\
			\hline
			\multirow{2}*{ 0.1 } & 0.5 (2e-04) & 0.51 (3e-04) & 0.53 (3e-04) & 0.55 (2e-04) & 0.62 (2e-04) & 0.74 (2e-04) & 0.9 (8e-05) & 1 (3e-06) & 1 (0e+00) & \multirow{2}*{\diagbox[dir=SW]{~~~~~~}{~~~~~~}} & 0 (0e+00) & \multirow{2}*{ 0.9 } \\ 
			& 0.5 (2e-04) & 0.51 (2e-04) & 0.53 (3e-04) & 0.56 (3e-04) & 0.63 (2e-04) & 0.74 (2e-04) & 0.9 (1e-04) & 0.99 (7e-06) & 1 (0e+00) & & 0.98 (2e-05)  \\ 
			\hline
			\multirow{2}*{ 0.2 } & 0.51 (2e-04) & 0.52 (3e-04) & 0.55 (3e-04) & 0.62 (2e-04) & 0.73 (2e-04) & 0.9 (9e-05) & 1 (4e-06) & 1 (0e+00) & \multirow{2}*{\diagbox[dir=SW]{~~~~~~}{~~~~~~}} & 0 (0e+00) & 0 (0e+00) & \multirow{2}*{ 0.8 } \\ 
			& 0.51 (2e-04) & 0.52 (3e-04) & 0.56 (2e-04) & 0.62 (3e-04) & 0.74 (2e-04) & 0.9 (1e-04) & 0.99 (5e-06) & 1 (0e+00) & & 1 (5e-08) & 0.78 (2e-04)  \\ 
			\hline
			\multirow{2}*{ 0.3 } & 0.51 (3e-04) & 0.54 (3e-04) & 0.6 (2e-04) & 0.72 (2e-04) & 0.89 (1e-04) & 1 (5e-06) & 1 (0e+00) & \multirow{2}*{\diagbox[dir=SW]{~~~~~~}{~~~~~~}} & 0 (0e+00) & 0 (0e+00) & 0 (0e+00) & \multirow{2}*{ 0.7 } \\ 
			& 0.51 (2e-04) & 0.55 (3e-04) & 0.61 (3e-04) & 0.73 (2e-04) & 0.89 (8e-05) & 0.99 (4e-06) & 1 (0e+00) & & 1 (0e+00) & 0.95 (6e-05) & 0.65 (3e-04)  \\ 
			\hline
			\multirow{2}*{ 0.4 } & 0.52 (3e-04) & 0.58 (2e-04) & 0.7 (2e-04) & 0.88 (1e-04) & 1 (3e-06) & 1 (0e+00) & \multirow{2}*{\diagbox[dir=SW]{~~~~~~}{~~~~~~}} & 0 (0e+00) & 0 (0e+00) & 0 (0e+00) & 0 (0e+00) & \multirow{2}*{ 0.6 } \\ 
			& 0.52 (3e-04) & 0.59 (2e-04) & 0.71 (2e-04) & 0.88 (9e-05) & 0.99 (5e-06) & 1 (0e+00) & & 1 (0e+00) & 0.98 (2e-05) & 0.8 (2e-04) & 0.58 (2e-04)  \\ 
			\hline
			\multirow{2}*{ 0.5 } & 0.54 (2e-04) & 0.66 (2e-04) & 0.86 (2e-04) & 0.99 (6e-06) & 1 (0e+00) & \multirow{2}*{\diagbox[dir=SW]{~~~~~~}{~~~~~~}} & 0 (0e+00) & 0 (0e+00) & 0 (0e+00) & 0 (0e+00) & 0 (0e+00) & \multirow{2}*{ 0.5 } \\ 
			& 0.54 (3e-04) & 0.67 (2e-04) & 0.86 (1e-04) & 0.99 (1e-05) & 1 (0e+00) & & 1 (0e+00) & 0.99 (8e-06) & 0.86 (1e-04) & 0.67 (2e-04) & 0.54 (2e-04)  \\ 
			\hline
			\multirow{2}*{ 0.6 } & 0.58 (2e-04) & 0.8 (2e-04) & 0.99 (1e-05) & 1 (0e+00) & \multirow{2}*{\diagbox[dir=SW]{~~~~~~}{~~~~~~}} & 0 (0e+00) & 0 (0e+00) & 0 (0e+00) & 0 (0e+00) & 0 (0e+00) & 0 (0e+00) & \multirow{2}*{ 0.4 } \\ 
			& 0.58 (3e-04) & 0.8 (1e-04) & 0.98 (1e-05) & 1 (0e+00) & & 1 (0e+00) & 0.99 (7e-06) & 0.88 (1e-04) & 0.71 (2e-04) & 0.58 (2e-04) & 0.52 (2e-04)  \\ 
			\hline
			\multirow{2}*{ 0.7 } & 0.65 (3e-04) & 0.96 (4e-05) & 1 (0e+00) & \multirow{2}*{\diagbox[dir=SW]{~~~~~~}{~~~~~~}} & 0 (0e+00) & 0 (0e+00) & 0 (0e+00) & 0 (0e+00) & 0 (0e+00) & 0 (0e+00) & 0 (0e+00) & \multirow{2}*{ 0.3 } \\ 
			& 0.65 (2e-04) & 0.95 (5e-05) & 1 (0e+00) & & 1 (0e+00) & 0.99 (5e-06) & 0.89 (1e-04) & 0.73 (2e-04) & 0.61 (2e-04) & 0.54 (3e-04) & 0.51 (3e-04)  \\ 
			\hline
			\multirow{2}*{ 0.8 } & 0.8 (1e-04) & 1 (0e+00) & \multirow{2}*{\diagbox[dir=SW]{~~~~~~}{~~~~~~}} & 0 (0e+00) & 0 (0e+00) & 0 (0e+00) & 0 (0e+00) & 0 (0e+00) & 0 (0e+00) & 0 (0e+00) & 0 (0e+00) & \multirow{2}*{ 0.2 } \\ 
			& 0.78 (2e-04) & 1 (1e-08) & & 1 (0e+00) & 0.99 (5e-06) & 0.9 (1e-04) & 0.74 (2e-04) & 0.62 (3e-04) & 0.56 (2e-04) & 0.52 (2e-04) & 0.51 (2e-04)  \\ 
			\hline
			\multirow{2}*{ 0.9 } & 1 (4e-06) & \multirow{2}*{\diagbox[dir=SW]{~~~~~~}{~~~~~~}} & 0 (0e+00) & 0 (0e+00) & 0 (0e+00) & 0 (0e+00) & 0 (0e+00) & 0 (0e+00) & 0 (0e+00) & 0 (0e+00) & 0 (0e+00) & \multirow{2}*{ 0.1 } \\ 
			& 0.98 (2e-05) & & 1 (0e+00) & 0.99 (5e-06) & 0.9 (9e-05) & 0.74 (2e-04) & 0.63 (3e-04) & 0.56 (2e-04) & 0.53 (3e-04) & 0.51 (3e-04) & 0.5 (2e-04)  \\ 
			\hline
			& \diagbox[dir=SW]{~~~~~~}{~~~~~~} & & 0.85 & 0.75 & 0.65 & 0.55 & 0.45 & 0.35 & 0.25 & 0.15 & 0.05 & \backslashbox{$\eta$}{$\rho$} \\
			\hline
			\bottomrule[2pt]
			\multicolumn{3}{c}{} & \multicolumn{10}{c}{\Large Different parameters} \\
		\end{tabular}
		\label{graph_simul_same_diff}
	}}
\end{table}

\subsection{Non-Gaussian Continuous Data}
The Copula function is adopted to generate the continuous data following non-Gaussian distribution. The data generation process with given correlation $\rho$ and target marginal distribution $f(x)$ is as follows:
\begin{itemize}
	\item[ ]
		\begin{itemize}
			\setlength{\itemsep}{0pt}
			\setlength{\parsep}{0pt}
			\setlength{\parskip}{0pt}
			\item[Step 2.1:] Generate the increments $(\Delta \boldsymbol{X}^*, \Delta \boldsymbol{Y}^*)$ based on Model~\eqref{degeneratedmodel} with given $(\mu^x,\mu^y), (\sigma^2_1,\sigma^2_2), \rho$;
			\item[Step 2.2:] For each increment $(\Delta x^*, \Delta y^*)$, conduct the following steps to get the target increments $(\Delta \boldsymbol{X}, \Delta \boldsymbol{Y})$:
				\begin{itemize}
					\setlength{\itemsep}{0pt}
					\setlength{\parsep}{0pt}
					\setlength{\parskip}{0pt}
					\item[Step 2.2.1:] Calculate its marginal cumulative probabilities: $\Phi_1=\Phi(\frac{\Delta x^*-\mu^x}{\sigma_1}), \Phi_2=\Phi(\frac{\Delta y^*-\mu^y}{\sigma_2})$;
					\item[Step 2.2.2:] Obtain the target sample $(\Delta x, \Delta y)$: $\Delta x=F^{-1}(\Phi_1), \Delta y=F^{-1}(\Phi_2)$, where $F(\cdot)$ is the cumulative distribution function corresponding to $f(\cdot)$;
				\end{itemize}
			\item[Step 2.3:] Given the root point $(X^0, Y^0)$, according to $(\Delta \boldsymbol{X}, \Delta \boldsymbol{Y})$, construct the first pair of tree-shaped datasets $(\boldsymbol{X}_1, \boldsymbol{Y}_1)$;
			\item[Step 2.4:] Do the same to generate the second pair of tree-shaped datasets $(\boldsymbol{X}_2, \boldsymbol{Y}_2)$.
		\end{itemize}
\end{itemize}

In this simulation, three marginal distributions $f(x)$ are setting as Gamma, F and Student-t distributions, and the results are displayed in Table~\ref{graph_simul_same_diff_gamma}, Table~\ref{graph_simul_same_diff_f} and Table~\ref{graph_simul_same_diff_t}, respectively. We can easily find that the proposed method can still achieve high accuracy even on non-Gaussian continuous data, which indicates the well generalization of the proposed method.
\begin{table}[!h]
	\centering
	\caption{Simulation results based on the Gamma distribution.}
	\resizebox{5.7in}{1.6in}{
	\resizebox{\linewidth}{!}{
		\begin{tabular}{c|c|c|c|c|c|c|c|c|c|c|c|c}
			\multicolumn{10}{c}{\Large Same parameters} & \multicolumn{3}{c}{} \\
			\hline
			\bottomrule[2pt]
			\backslashbox{$\rho$}{$\eta$} & 0.05 & 0.15 & 0.25 & 0.35 & 0.45 & 0.55 & 0.65 & 0.75 & 0.85 & & \diagbox[dir=SW]{~~~~~~}{~~~~~~} & \\
			\hline
			\multirow{2}*{ 0.1 } & 0.5 (3e-04) & 0.5 (3e-04) & 0.52 (2e-04) & 0.55 (2e-04) & 0.61 (3e-04) & 0.74 (2e-04) & 0.92 (7e-05) & 1 (2e-06) & 1 (0e+00) & \multirow{2}*{\diagbox[dir=SW]{~~~~~~}{~~~~~~}} & 0.08 (7e-05) & \multirow{2}*{ 0.9 } \\ 
			& 0.5 (3e-04) & 0.51 (3e-04) & 0.52 (2e-04) & 0.56 (2e-04) & 0.64 (2e-04) & 0.77 (2e-04) & 0.93 (5e-05) & 1 (2e-06) & 1 (0e+00) & & 0.99 (1e-05)  \\ 
			\hline
			\multirow{2}*{ 0.2 } & 0.5 (2e-04) & 0.51 (2e-04) & 0.54 (3e-04) & 0.61 (2e-04) & 0.74 (2e-04) & 0.91 (8e-05) & 1 (2e-06) & 1 (0e+00) & \multirow{2}*{\diagbox[dir=SW]{~~~~~~}{~~~~~~}} & 0.69 (2e-04) & 0 (1e-07) & \multirow{2}*{ 0.8 } \\ 
			& 0.5 (2e-04) & 0.52 (2e-04) & 0.56 (3e-04) & 0.63 (2e-04) & 0.77 (2e-04) & 0.93 (6e-05) & 1 (2e-06) & 1 (0e+00) & & 1 (3e-08) & 0.8 (1e-04)  \\ 
			\hline
			\multirow{2}*{ 0.3 } & 0.51 (3e-04) & 0.54 (3e-04) & 0.6 (3e-04) & 0.74 (2e-04) & 0.91 (7e-05) & 1 (2e-06) & 1 (0e+00) & \multirow{2}*{\diagbox[dir=SW]{~~~~~~}{~~~~~~}} & 0.91 (8e-05) & 0 (3e-06) & 0 (0e+00) & \multirow{2}*{ 0.7 } \\ 
			& 0.51 (3e-04) & 0.54 (3e-04) & 0.62 (3e-04) & 0.76 (2e-04) & 0.93 (6e-05) & 1 (2e-06) & 1 (0e+00) & & 1 (0e+00) & 0.97 (3e-05) & 0.67 (2e-04)  \\ 
			\hline
			\multirow{2}*{ 0.4 } & 0.52 (2e-04) & 0.59 (2e-04) & 0.72 (2e-04) & 0.9 (1e-04) & 1 (3e-06) & 1 (0e+00) & \multirow{2}*{\diagbox[dir=SW]{~~~~~~}{~~~~~~}} & 0.96 (3e-05) & 0.01 (1e-05) & 0 (4e-08) & 0 (0e+00) & \multirow{2}*{ 0.6 } \\ 
			& 0.52 (3e-04) & 0.6 (2e-04) & 0.74 (2e-04) & 0.92 (9e-05) & 1 (2e-06) & 1 (0e+00) & & 1 (0e+00) & 0.99 (7e-06) & 0.83 (2e-04) & 0.59 (2e-04)  \\ 
			\hline
			\multirow{2}*{ 0.5 } & 0.54 (3e-04) & 0.68 (2e-04) & 0.88 (1e-04) & 1 (3e-06) & 1 (0e+00) & \multirow{2}*{\diagbox[dir=SW]{~~~~~~}{~~~~~~}} & 0.98 (2e-05) & 0.02 (2e-05) & 0 (6e-08) & 0 (0e+00) & 0 (0e+00) & \multirow{2}*{ 0.5 } \\ 
			& 0.54 (3e-04) & 0.69 (2e-04) & 0.89 (1e-04) & 1 (2e-06) & 1 (0e+00) & & 1 (0e+00) & 1 (4e-06) & 0.89 (1e-04) & 0.69 (2e-04) & 0.55 (2e-04)  \\ 
			\hline
			\multirow{2}*{ 0.6 } & 0.58 (2e-04) & 0.82 (1e-04) & 0.99 (6e-06) & 1 (0e+00) & \multirow{2}*{\diagbox[dir=SW]{~~~~~~}{~~~~~~}} & 0.98 (2e-05) & 0.02 (2e-05) & 0 (8e-08) & 0 (0e+00) & 0 (0e+00) & 0 (0e+00) & \multirow{2}*{ 0.4 } \\ 
			& 0.59 (2e-04) & 0.83 (1e-04) & 0.99 (5e-06) & 1 (0e+00) & & 1 (0e+00) & 1 (2e-06) & 0.92 (1e-04) & 0.74 (2e-04) & 0.6 (3e-04) & 0.52 (2e-04)  \\ 
			\hline
			\multirow{2}*{ 0.7 } & 0.66 (2e-04) & 0.97 (3e-05) & 1 (0e+00) & \multirow{2}*{\diagbox[dir=SW]{~~~~~~}{~~~~~~}} & 0.98 (2e-05) & 0.02 (2e-05) & 0 (8e-08) & 0 (0e+00) & 0 (0e+00) & 0 (0e+00) & 0 (0e+00) & \multirow{2}*{ 0.3 } \\ 
			& 0.67 (2e-04) & 0.97 (3e-05) & 1 (0e+00) & & 1 (0e+00) & 1 (2e-06) & 0.93 (7e-05) & 0.76 (2e-04) & 0.62 (2e-04) & 0.55 (3e-04) & 0.51 (2e-04)  \\ 
			\hline
			\multirow{2}*{ 0.8 } & 0.8 (2e-04) & 1 (2e-08) & \multirow{2}*{\diagbox[dir=SW]{~~~~~~}{~~~~~~}} & 0.98 (2e-05) & 0.02 (1e-05) & 0 (4e-08) & 0 (0e+00) & 0 (0e+00) & 0 (0e+00) & 0 (0e+00) & 0 (0e+00) & \multirow{2}*{ 0.2 } \\ 
			& 0.8 (2e-04) & 1 (3e-08) & & 1 (0e+00) & 1 (2e-06) & 0.93 (6e-05) & 0.77 (2e-04) & 0.64 (2e-04) & 0.56 (2e-04) & 0.52 (2e-04) & 0.51 (3e-04)  \\ 
			\hline
			\multirow{2}*{ 0.9 } & 0.99 (1e-05) & \multirow{2}*{\diagbox[dir=SW]{~~~~~~}{~~~~~~}} & 0.98 (2e-05) & 0.02 (2e-05) & 0 (9e-08) & 0 (0e+00) & 0 (0e+00) & 0 (0e+00) & 0 (0e+00) & 0 (0e+00) & 0 (0e+00) & \multirow{2}*{ 0.1 } \\ 
			& 0.99 (2e-05) & & 1 (0e+00) & 1 (2e-06) & 0.93 (8e-05) & 0.77 (2e-04) & 0.64 (2e-04) & 0.56 (2e-04) & 0.53 (2e-04) & 0.51 (2e-04) & 0.5 (2e-04)  \\ 
			\hline
			& \diagbox[dir=SW]{~~~~~~}{~~~~~~} & & 0.85 & 0.75 & 0.65 & 0.55 & 0.45 & 0.35 & 0.25 & 0.15 & 0.05 & \backslashbox{$\eta$}{$\rho$} \\
			\hline
			\bottomrule[2pt]
			\multicolumn{3}{c}{} & \multicolumn{10}{c}{\Large Different parameters} \\
		\end{tabular}
		\label{graph_simul_same_diff_gamma}
	}}
\end{table}
\begin{table}[!h]
	\vspace{-0.8cm}
	\centering
	\caption{Simulation results based on the F distribution.}
	\resizebox{5.7in}{1.6in}{
		\resizebox{\linewidth}{!}{
			\begin{tabular}{c|c|c|c|c|c|c|c|c|c|c|c|c}
				\multicolumn{10}{c}{\Large Same parameters} & \multicolumn{3}{c}{} \\
				\hline
				\bottomrule[2pt]
				\backslashbox{$\rho$}{$\eta$} & 0.05 & 0.15 & 0.25 & 0.35 & 0.45 & 0.55 & 0.65 & 0.75 & 0.85 & & \diagbox[dir=SW]{~~~~~~}{~~~~~~} & \\
				\hline
				\multirow{2}*{ 0.1 } & 0.5 (3e-04) & 0.5 (3e-04) & 0.52 (3e-04) & 0.55 (2e-04) & 0.61 (2e-04) & 0.74 (2e-04) & 0.92 (6e-05) & 1 (2e-06) & 1 (0e+00) & \multirow{2}*{\diagbox[dir=SW]{~~~~~~}{~~~~~~}} & 0 (0e+00) & \multirow{2}*{ 0.9 } \\ 
				& 0.5 (3e-04) & 0.51 (3e-04) & 0.52 (2e-04) & 0.56 (2e-04) & 0.63 (2e-04) & 0.76 (2e-04) & 0.92 (7e-05) & 1 (3e-06) & 1 (0e+00) & & 0.92 (6e-05)  \\ 
				\hline
				\multirow{2}*{ 0.2 } & 0.5 (3e-04) & 0.51 (2e-04) & 0.54 (3e-04) & 0.61 (2e-04) & 0.74 (2e-04) & 0.91 (7e-05) & 1 (2e-06) & 1 (0e+00) & \multirow{2}*{\diagbox[dir=SW]{~~~~~~}{~~~~~~}} & 0 (4e-08) & 0 (0e+00) & \multirow{2}*{ 0.8 } \\ 
				& 0.5 (2e-04) & 0.52 (2e-04) & 0.55 (3e-04) & 0.63 (2e-04) & 0.75 (2e-04) & 0.92 (9e-05) & 1 (3e-06) & 1 (0e+00) & & 1 (6e-07) & 0.61 (3e-04)  \\ 
				\hline
				\multirow{2}*{ 0.3 } & 0.51 (3e-04) & 0.54 (3e-04) & 0.6 (3e-04) & 0.74 (2e-04) & 0.91 (7e-05) & 1 (2e-06) & 1 (0e+00) & \multirow{2}*{\diagbox[dir=SW]{~~~~~~}{~~~~~~}} & 0 (7e-07) & 0 (0e+00) & 0 (0e+00) & \multirow{2}*{ 0.7 } \\ 
				& 0.51 (3e-04) & 0.54 (2e-04) & 0.61 (3e-04) & 0.75 (2e-04) & 0.91 (7e-05) & 1 (3e-06) & 1 (0e+00) & & 1 (2e-08) & 0.88 (1e-04) & 0.47 (2e-04)  \\ 
				\hline
				\multirow{2}*{ 0.4 } & 0.52 (3e-04) & 0.59 (2e-04) & 0.72 (2e-04) & 0.9 (9e-05) & 1 (3e-06) & 1 (0e+00) & \multirow{2}*{\diagbox[dir=SW]{~~~~~~}{~~~~~~}} & 0 (2e-06) & 0 (0e+00) & 0 (0e+00) & 0 (0e+00) & \multirow{2}*{ 0.6 } \\ 
				& 0.52 (3e-04) & 0.59 (2e-04) & 0.73 (2e-04) & 0.9 (1e-04) & 1 (3e-06) & 1 (0e+00) & & 1 (1e-08) & 0.94 (5e-05) & 0.65 (2e-04) & 0.4 (2e-04)  \\ 
				\hline
				\multirow{2}*{ 0.5 } & 0.54 (2e-04) & 0.68 (2e-04) & 0.88 (1e-04) & 1 (3e-06) & 1 (0e+00) & \multirow{2}*{\diagbox[dir=SW]{~~~~~~}{~~~~~~}} & 0 (5e-06) & 0 (0e+00) & 0 (0e+00) & 0 (0e+00) & 0 (0e+00) & \multirow{2}*{ 0.5 } \\ 
				& 0.54 (2e-04) & 0.68 (2e-04) & 0.88 (1e-04) & 0.99 (6e-06) & 1 (0e+00) & & 1 (0e+00) & 0.97 (3e-05) & 0.73 (2e-04) & 0.5 (3e-04) & 0.37 (2e-04)  \\ 
				\hline
				\multirow{2}*{ 0.6 } & 0.58 (2e-04) & 0.82 (1e-04) & 0.99 (6e-06) & 1 (0e+00) & \multirow{2}*{\diagbox[dir=SW]{~~~~~~}{~~~~~~}} & 0 (4e-06) & 0 (0e+00) & 0 (0e+00) & 0 (0e+00) & 0 (0e+00) & 0 (0e+00) & \multirow{2}*{ 0.4 } \\ 
				& 0.58 (3e-04) & 0.82 (1e-04) & 0.99 (8e-06) & 1 (0e+00) & & 1 (0e+00) & 0.97 (2e-05) & 0.77 (2e-04) & 0.54 (3e-04) & 0.42 (3e-04) & 0.36 (2e-04)  \\ 
				\hline
				\multirow{2}*{ 0.7 } & 0.66 (3e-04) & 0.97 (3e-05) & 1 (0e+00) & \multirow{2}*{\diagbox[dir=SW]{~~~~~~}{~~~~~~}} & 0.01 (5e-06) & 0 (0e+00) & 0 (0e+00) & 0 (0e+00) & 0 (0e+00) & 0 (0e+00) & 0 (0e+00) & \multirow{2}*{ 0.3 } \\ 
				& 0.66 (2e-04) & 0.96 (3e-05) & 1 (0e+00) & & 1 (0e+00) & 0.98 (2e-05) & 0.78 (2e-04) & 0.57 (3e-04) & 0.44 (3e-04) & 0.38 (2e-04) & 0.35 (3e-04)  \\ 
				\hline
				\multirow{2}*{ 0.8 } & 0.8 (2e-04) & 1 (2e-08) & \multirow{2}*{\diagbox[dir=SW]{~~~~~~}{~~~~~~}} & 0.01 (7e-06) & 0 (0e+00) & 0 (0e+00) & 0 (0e+00) & 0 (0e+00) & 0 (0e+00) & 0 (0e+00) & 0 (0e+00) & \multirow{2}*{ 0.2 } \\ 
				& 0.79 (2e-04) & 1 (3e-08) & & 1 (0e+00) & 0.98 (3e-05) & 0.79 (1e-04) & 0.57 (2e-04) & 0.45 (2e-04) & 0.38 (2e-04) & 0.36 (3e-04) & 0.35 (2e-04)  \\ 
				\hline
				\multirow{2}*{ 0.9 } & 0.99 (1e-05) & \multirow{2}*{\diagbox[dir=SW]{~~~~~~}{~~~~~~}} & 0.01 (5e-06) & 0 (0e+00) & 0 (0e+00) & 0 (0e+00) & 0 (0e+00) & 0 (0e+00) & 0 (0e+00) & 0 (0e+00) & 0 (0e+00) & \multirow{2}*{ 0.1 } \\ 
				& 0.98 (2e-05) & & 1 (0e+00) & 0.98 (2e-05) & 0.79 (2e-04) & 0.58 (2e-04) & 0.45 (2e-04) & 0.39 (2e-04) & 0.36 (2e-04) & 0.35 (2e-04) & 0.35 (2e-04)  \\ 
				\hline
				& \diagbox[dir=SW]{~~~~~~}{~~~~~~} & & 0.85 & 0.75 & 0.65 & 0.55 & 0.45 & 0.35 & 0.25 & 0.15 & 0.05 & \backslashbox{$\eta$}{$\rho$} \\
				\hline
				\bottomrule[2pt]
				\multicolumn{3}{c}{} & \multicolumn{10}{c}{\Large Different parameters} \\
			\end{tabular}
			\label{graph_simul_same_diff_f}
		}}
\end{table}
\begin{table}[!h]
	\centering
	\caption{Simulation results based on the Student-t distribution.}
	\resizebox{5.7in}{1.6in}{
		\resizebox{\linewidth}{!}{
			\begin{tabular}{c|c|c|c|c|c|c|c|c|c|c|c|c}
				\multicolumn{10}{c}{\Large Same parameters} & \multicolumn{3}{c}{} \\
				\hline
				\bottomrule[2pt]
				\backslashbox{$\rho$}{$\eta$} & 0.05 & 0.15 & 0.25 & 0.35 & 0.45 & 0.55 & 0.65 & 0.75 & 0.85 & & \diagbox[dir=SW]{~~~~~~}{~~~~~~} & \\
				\hline
				\multirow{2}*{ 0.1 } & 0.5 (3e-04) & 0.5 (3e-04) & 0.51 (3e-04) & 0.54 (2e-04) & 0.61 (3e-04) & 0.73 (2e-04) & 0.9 (8e-05) & 1 (5e-06) & 1 (0e+00) & \multirow{2}*{\diagbox[dir=SW]{~~~~~~}{~~~~~~}} & 0 (0e+00) & \multirow{2}*{ 0.9 } \\ 
				& 0.5 (2e-04) & 0.51 (3e-04) & 0.52 (2e-04) & 0.56 (2e-04) & 0.64 (2e-04) & 0.77 (1e-04) & 0.93 (5e-05) & 1 (2e-06) & 1 (0e+00) & & 0.99 (1e-05)  \\ 
				\hline
				\multirow{2}*{ 0.2 } & 0.5 (3e-04) & 0.51 (2e-04) & 0.54 (2e-04) & 0.61 (3e-04) & 0.73 (2e-04) & 0.9 (9e-05) & 1 (3e-06) & 1 (0e+00) & \multirow{2}*{\diagbox[dir=SW]{~~~~~~}{~~~~~~}} & 0 (4e-08) & 0 (0e+00) & \multirow{2}*{ 0.8 } \\ 
				& 0.5 (2e-04) & 0.52 (2e-04) & 0.56 (3e-04) & 0.63 (2e-04) & 0.77 (2e-04) & 0.93 (6e-05) & 1 (2e-06) & 1 (0e+00) & & 1 (3e-08) & 0.8 (1e-04)  \\ 
				\hline
				\multirow{2}*{ 0.3 } & 0.51 (3e-04) & 0.53 (2e-04) & 0.6 (3e-04) & 0.72 (2e-04) & 0.9 (8e-05) & 1 (5e-06) & 1 (0e+00) & \multirow{2}*{\diagbox[dir=SW]{~~~~~~}{~~~~~~}} & 0 (4e-07) & 0 (0e+00) & 0 (0e+00) & \multirow{2}*{ 0.7 } \\ 
				& 0.51 (3e-04) & 0.54 (3e-04) & 0.62 (3e-04) & 0.76 (2e-04) & 0.93 (6e-05) & 1 (2e-06) & 1 (0e+00) & & 1 (0e+00) & 0.97 (3e-05) & 0.66 (2e-04)  \\ 
				\hline
				\multirow{2}*{ 0.4 } & 0.52 (3e-04) & 0.58 (2e-04) & 0.71 (2e-04) & 0.89 (1e-04) & 1 (4e-06) & 1 (0e+00) & \multirow{2}*{\diagbox[dir=SW]{~~~~~~}{~~~~~~}} & 0 (9e-07) & 0 (0e+00) & 0 (0e+00) & 0 (0e+00) & \multirow{2}*{ 0.6 } \\ 
				& 0.52 (3e-04) & 0.6 (2e-04) & 0.74 (2e-04) & 0.91 (8e-05) & 1 (2e-06) & 1 (0e+00) & & 1 (0e+00) & 0.99 (8e-06) & 0.83 (2e-04) & 0.59 (2e-04)  \\ 
				\hline
				\multirow{2}*{ 0.5 } & 0.54 (3e-04) & 0.67 (2e-04) & 0.87 (1e-04) & 0.99 (6e-06) & 1 (0e+00) & \multirow{2}*{\diagbox[dir=SW]{~~~~~~}{~~~~~~}} & 0 (2e-06) & 0 (0e+00) & 0 (0e+00) & 0 (0e+00) & 0 (0e+00) & \multirow{2}*{ 0.5 } \\ 
				& 0.55 (2e-04) & 0.69 (2e-04) & 0.89 (1e-04) & 1 (2e-06) & 1 (0e+00) & & 1 (0e+00) & 1 (4e-06) & 0.89 (1e-04) & 0.69 (2e-04) & 0.55 (2e-04)  \\ 
				\hline
				\multirow{2}*{ 0.6 } & 0.58 (3e-04) & 0.81 (2e-04) & 0.99 (1e-05) & 1 (0e+00) & \multirow{2}*{\diagbox[dir=SW]{~~~~~~}{~~~~~~}} & 0 (2e-06) & 0 (0e+00) & 0 (0e+00) & 0 (0e+00) & 0 (0e+00) & 0 (0e+00) & \multirow{2}*{ 0.4 } \\ 
				& 0.59 (3e-04) & 0.83 (1e-04) & 0.99 (5e-06) & 1 (0e+00) & & 1 (0e+00) & 1 (2e-06) & 0.91 (9e-05) & 0.74 (2e-04) & 0.6 (3e-04) & 0.52 (2e-04)  \\ 
				\hline
				\multirow{2}*{ 0.7 } & 0.65 (2e-04) & 0.96 (5e-05) & 1 (0e+00) & \multirow{2}*{\diagbox[dir=SW]{~~~~~~}{~~~~~~}} & 0 (2e-06) & 0 (0e+00) & 0 (0e+00) & 0 (0e+00) & 0 (0e+00) & 0 (0e+00) & 0 (0e+00) & \multirow{2}*{ 0.3 } \\ 
				& 0.67 (2e-04) & 0.97 (3e-05) & 1 (0e+00) & & 1 (0e+00) & 1 (2e-06) & 0.93 (7e-05) & 0.76 (2e-04) & 0.62 (2e-04) & 0.55 (3e-04) & 0.51 (2e-04)  \\ 
				\hline
				\multirow{2}*{ 0.8 } & 0.79 (2e-04) & 1 (2e-08) & \multirow{2}*{\diagbox[dir=SW]{~~~~~~}{~~~~~~}} & 0 (3e-06) & 0 (0e+00) & 0 (0e+00) & 0 (0e+00) & 0 (0e+00) & 0 (0e+00) & 0 (0e+00) & 0 (0e+00) & \multirow{2}*{ 0.2 } \\ 
				& 0.8 (2e-04) & 1 (3e-08) & & 1 (0e+00) & 1 (3e-06) & 0.93 (5e-05) & 0.77 (2e-04) & 0.63 (2e-04) & 0.55 (2e-04) & 0.52 (2e-04) & 0.51 (3e-04)  \\ 
				\hline
				\multirow{2}*{ 0.9 } & 0.99 (1e-05) & \multirow{2}*{\diagbox[dir=SW]{~~~~~~}{~~~~~~}} & 0 (3e-06) & 0 (0e+00) & 0 (0e+00) & 0 (0e+00) & 0 (0e+00) & 0 (0e+00) & 0 (0e+00) & 0 (0e+00) & 0 (0e+00) & \multirow{2}*{ 0.1 } \\ 
				& 0.99 (2e-05) & & 1 (0e+00) & 1 (2e-06) & 0.93 (8e-05) & 0.77 (2e-04) & 0.64 (2e-04) & 0.56 (2e-04) & 0.53 (2e-04) & 0.51 (3e-04) & 0.5 (3e-04)  \\ 
				\hline
				& \diagbox[dir=SW]{~~~~~~}{~~~~~~} & & 0.85 & 0.75 & 0.65 & 0.55 & 0.45 & 0.35 & 0.25 & 0.15 & 0.05 & \backslashbox{$\eta$}{$\rho$} \\
				\hline
				\bottomrule[2pt]
				\multicolumn{3}{c}{} & \multicolumn{10}{c}{\Large Different parameters} \\
			\end{tabular}
			\label{graph_simul_same_diff_t}
		}}
\end{table}

\subsection{Discrete Data}
Besides continuous data, discrete data with specified correlation $\rho$ is generated by three discretization methods.

The first approach uses the Copula function based on the Poisson distribution to generate discrete data and the detailed steps are as follows:
\begin{itemize}
	\item[ ]
	\begin{itemize}
		\setlength{\itemsep}{0pt}
		\setlength{\parsep}{0pt}
		\setlength{\parskip}{0pt}
		\item[Step 2.1:] Generate the increments $(\Delta \boldsymbol{X}^*, \Delta \boldsymbol{Y}^*)$ based on Model~\eqref{degeneratedmodel} with given $(\mu^x,\mu^y), (\sigma^2_1,\sigma^2_2), \rho$;
		\item[Step 2.2:] For each increment $(\Delta x^*, \Delta y^*)$, conduct the following steps to get the target increments $(\Delta \boldsymbol{X}, \Delta \boldsymbol{Y})$:
			\begin{itemize}
				\setlength{\itemsep}{0pt}
				\setlength{\parsep}{0pt}
				\setlength{\parskip}{0pt}
				\item[Step 2.2.1:] Calculate its marginal cumulative probabilities: $\Phi_1=\Phi(\frac{\Delta x^*-\mu^x}{\sigma_1}), \Phi_2=\Phi(\frac{\Delta y^*-\mu^y}{\sigma_2})$;
				\item[Step 2.2.2:] Obtain the target sample $(\Delta x, \Delta y)$: $\Delta x=\arg_x \{ F(x) \leq \Phi_1 < F(x+1) \}$, $\Delta y=\arg_y \{ F(y) \leq \Phi_2 < F(y+1) \}$, where $F(\cdot)$ is the cumulative mass function corresponding to $f(\cdot)$;
			\end{itemize}
		\item[Step 2.3:] Given the root point $(X^0, Y^0)$, according to $(\Delta \boldsymbol{X}, \Delta \boldsymbol{Y})$, construct the first pair of tree-shaped datasets $(\boldsymbol{X}_1, \boldsymbol{Y}_1)$;
		\item[Step 2.4:] Do the same to generate the second pair of tree-shaped datasets $(\boldsymbol{X}_2, \boldsymbol{Y}_2)$.
	\end{itemize}
\end{itemize}

Other two approaches are two well-known unsupervised discretization methods: equal-width discretization and equal-frequency discretization \citep{2005Discretization}. Equal-width discretization divides the range of the attribute into a fixed number of intervals with equal length. Equal-frequency discretization also has a fixed number of intervals, but the intervals are chosen so that each one has the same or approximately the same number of samples. Synthesized incremental datasets generated by Step 2.1 are discretized for two dimensions independently, and the index of interval, where $\Delta x^*$ or $\Delta y^*$ is located, is regarded as the new sample after discretization, i.e., $\Delta x$ or $\Delta x$.

The results with and without normalization of these three methods are displayed in Table~\ref{graph_simul_same_diff_discrete_poisson}, Table~\ref{graph_simul_same_diff_discrete_dist} and Table~\ref{graph_simul_same_diff_discrete_freq}, respectively. We can easily know that the proposed method also performs well on discrete data.
\begin{table}[!h]
	\centering
	\caption{Simulation results based on the Poisson distribution.}
	\resizebox{5.7in}{1.6in}{
		\resizebox{\linewidth}{!}{
			\begin{tabular}{c|c|c|c|c|c|c|c|c|c|c|c|c}
				\multicolumn{10}{c}{\Large Same parameters} & \multicolumn{3}{c}{} \\
				\hline
				\bottomrule[2pt]
				\backslashbox{$\rho$}{$\eta$} & 0.05 & 0.15 & 0.25 & 0.35 & 0.45 & 0.55 & 0.65 & 0.75 & 0.85 & & \diagbox[dir=SW]{~~~~~~}{~~~~~~} & \\
				\hline
				\multirow{2}*{ 0.1 } & 0.5 (3e-04) & 0.51 (3e-04) & 0.51 (2e-04) & 0.54 (2e-04) & 0.59 (2e-04) & 0.71 (2e-04) & 0.88 (9e-05) & 0.99 (1e-05) & 1 (0e+00) & \multirow{2}*{\diagbox[dir=SW]{~~~~~~}{~~~~~~}} & 0.14 (1e-04) & \multirow{2}*{ 0.9 } \\ 
				& 0.5 (2e-04) & 0.51 (3e-04) & 0.53 (3e-04) & 0.56 (2e-04) & 0.63 (2e-04) & 0.76 (2e-04) & 0.92 (7e-05) & 1 (2e-06) & 1 (0e+00) & & 0.98 (2e-05)  \\ 
				\hline
				\multirow{2}*{ 0.2 } & 0.5 (3e-04) & 0.51 (3e-04) & 0.54 (2e-04) & 0.59 (2e-04) & 0.71 (2e-04) & 0.87 (9e-05) & 0.99 (7e-06) & 1 (0e+00) & \multirow{2}*{\diagbox[dir=SW]{~~~~~~}{~~~~~~}} & 0.76 (2e-04) & 0 (2e-06) & \multirow{2}*{ 0.8 } \\ 
				& 0.5 (3e-04) & 0.52 (3e-04) & 0.56 (2e-04) & 0.63 (3e-04) & 0.76 (2e-04) & 0.92 (8e-05) & 1 (2e-06) & 1 (0e+00) & & 1 (4e-08) & 0.8 (2e-04)  \\ 
				\hline
				\multirow{2}*{ 0.3 } & 0.51 (2e-04) & 0.53 (2e-04) & 0.59 (2e-04) & 0.7 (2e-04) & 0.87 (1e-04) & 0.99 (1e-05) & 1 (0e+00) & \multirow{2}*{\diagbox[dir=SW]{~~~~~~}{~~~~~~}} & 0.93 (6e-05) & 0.02 (2e-05) & 0 (1e-07) & \multirow{2}*{ 0.7 } \\ 
				& 0.51 (3e-04) & 0.54 (2e-04) & 0.62 (2e-04) & 0.75 (2e-04) & 0.92 (9e-05) & 1 (2e-06) & 1 (0e+00) & & 1 (0e+00) & 0.97 (4e-05) & 0.66 (2e-04)  \\ 
				\hline
				\multirow{2}*{ 0.4 } & 0.52 (3e-04) & 0.58 (2e-04) & 0.69 (2e-04) & 0.86 (1e-04) & 0.99 (1e-05) & 1 (0e+00) & \multirow{2}*{\diagbox[dir=SW]{~~~~~~}{~~~~~~}} & 0.97 (3e-05) & 0.04 (4e-05) & 0 (6e-07) & 0 (5e-08) & \multirow{2}*{ 0.6 } \\ 
				& 0.53 (3e-04) & 0.6 (3e-04) & 0.73 (2e-04) & 0.91 (7e-05) & 1 (3e-06) & 1 (0e+00) & & 1 (0e+00) & 0.99 (8e-06) & 0.83 (1e-04) & 0.59 (2e-04)  \\ 
				\hline
				\multirow{2}*{ 0.5 } & 0.54 (2e-04) & 0.65 (2e-04) & 0.84 (1e-04) & 0.99 (1e-05) & 1 (0e+00) & \multirow{2}*{\diagbox[dir=SW]{~~~~~~}{~~~~~~}} & 0.98 (1e-05) & 0.06 (5e-05) & 0 (1e-06) & 0 (7e-08) & 0 (2e-08) & \multirow{2}*{ 0.5 } \\ 
				& 0.54 (2e-04) & 0.69 (2e-04) & 0.89 (1e-04) & 1 (4e-06) & 1 (0e+00) & & 1 (0e+00) & 1 (4e-06) & 0.89 (9e-05) & 0.69 (2e-04) & 0.54 (2e-04)  \\ 
				\hline
				\multirow{2}*{ 0.6 } & 0.57 (3e-04) & 0.78 (1e-04) & 0.98 (2e-05) & 1 (0e+00) & \multirow{2}*{\diagbox[dir=SW]{~~~~~~}{~~~~~~}} & 0.98 (2e-05) & 0.07 (8e-05) & 0 (1e-06) & 0 (7e-08) & 0 (3e-08) & 0 (1e-08) & \multirow{2}*{ 0.4 } \\ 
				& 0.59 (3e-04) & 0.83 (1e-04) & 0.99 (9e-06) & 1 (0e+00) & & 1 (0e+00) & 1 (3e-06) & 0.91 (7e-05) & 0.74 (2e-04) & 0.6 (2e-04) & 0.52 (3e-04)  \\ 
				\hline
				\multirow{2}*{ 0.7 } & 0.64 (2e-04) & 0.94 (6e-05) & 1 (0e+00) & \multirow{2}*{\diagbox[dir=SW]{~~~~~~}{~~~~~~}} & 0.99 (1e-05) & 0.08 (8e-05) & 0 (2e-06) & 0 (1e-07) & 0 (3e-08) & 0 (1e-08) & 0 (1e-08) & \multirow{2}*{ 0.3 } \\ 
				& 0.66 (3e-04) & 0.97 (3e-05) & 1 (0e+00) & & 1 (0e+00) & 1 (2e-06) & 0.92 (8e-05) & 0.75 (2e-04) & 0.62 (3e-04) & 0.55 (3e-04) & 0.51 (3e-04)  \\ 
				\hline
				\multirow{2}*{ 0.8 } & 0.77 (2e-04) & 1 (5e-08) & \multirow{2}*{\diagbox[dir=SW]{~~~~~~}{~~~~~~}} & 0.99 (1e-05) & 0.08 (7e-05) & 0 (2e-06) & 0 (1e-07) & 0 (5e-08) & 0 (1e-08) & 0 (2e-08) & 0 (3e-08) & \multirow{2}*{ 0.2 } \\ 
				& 0.8 (2e-04) & 1 (1e-08) & & 1 (0e+00) & 1 (2e-06) & 0.93 (6e-05) & 0.76 (2e-04) & 0.63 (3e-04) & 0.55 (3e-04) & 0.52 (2e-04) & 0.51 (3e-04)  \\ 
				\hline
				\multirow{2}*{ 0.9 } & 0.98 (2e-05) & \multirow{2}*{\diagbox[dir=SW]{~~~~~~}{~~~~~~}} & 0.99 (1e-05) & 0.08 (8e-05) & 0 (1e-06) & 0 (1e-07) & 0 (2e-08) & 0 (1e-08) & 0 (0e+00) & 0 (0e+00) & 0 (1e-08) & \multirow{2}*{ 0.1 } \\ 
				& 0.98 (2e-05) & & 1 (0e+00) & 1 (2e-06) & 0.93 (6e-05) & 0.76 (2e-04) & 0.64 (3e-04) & 0.56 (3e-04) & 0.52 (2e-04) & 0.51 (2e-04) & 0.5 (3e-04)  \\ 
				\hline
				& \diagbox[dir=SW]{~~~~~~}{~~~~~~} & & 0.85 & 0.75 & 0.65 & 0.55 & 0.45 & 0.35 & 0.25 & 0.15 & 0.05 & \backslashbox{$\eta$}{$\rho$} \\
				\hline
				\bottomrule[2pt]
				\multicolumn{3}{c}{} & \multicolumn{10}{c}{\Large Different parameters} \\
			\end{tabular}
			\label{graph_simul_same_diff_discrete_poisson}
		}}
\end{table}
\begin{table}[!h]
	\vspace{-0.8cm}
	\centering
	\caption{Simulation results based on the distance criterion.}
	\resizebox{5.7in}{1.6in}{
		\resizebox{\linewidth}{!}{
			\begin{tabular}{c|c|c|c|c|c|c|c|c|c|c|c|c}
				\multicolumn{10}{c}{\Large Same parameters} & \multicolumn{3}{c}{} \\
				\hline
				\bottomrule[2pt]
				\backslashbox{$\rho$}{$\eta$} & 0.05 & 0.15 & 0.25 & 0.35 & 0.45 & 0.55 & 0.65 & 0.75 & 0.85 & & \diagbox[dir=SW]{~~~~~~}{~~~~~~} & \\
				\hline
				\multirow{2}*{ 0.1 } & 0.51 (2e-04) & 0.52 (2e-04) & 0.55 (2e-04) & 0.59 (3e-04) & 0.65 (2e-04) & 0.75 (2e-04) & 0.88 (1e-04) & 0.98 (2e-05) & 1 (4e-08) & \multirow{2}*{\diagbox[dir=SW]{~~~~~~}{~~~~~~}} & 0.6 (3e-04) & \multirow{2}*{ 0.9 } \\ 
				& 0.51 (3e-04) & 0.53 (2e-04) & 0.56 (3e-04) & 0.6 (3e-04) & 0.66 (2e-04) & 0.75 (2e-04) & 0.86 (1e-04) & 0.96 (3e-05) & 1 (5e-07) & & 0.89 (1e-04)  \\ 
				\hline
				\multirow{2}*{ 0.2 } & 0.51 (2e-04) & 0.53 (3e-04) & 0.57 (3e-04) & 0.64 (2e-04) & 0.74 (2e-04) & 0.87 (1e-04) & 0.98 (3e-05) & 1 (4e-08) & \multirow{2}*{\diagbox[dir=SW]{~~~~~~}{~~~~~~}} & 0.91 (9e-05) & 0.24 (2e-04) & \multirow{2}*{ 0.8 } \\ 
				& 0.51 (3e-04) & 0.54 (2e-04) & 0.58 (2e-04) & 0.64 (3e-04) & 0.73 (2e-04) & 0.85 (2e-04) & 0.96 (4e-05) & 1 (4e-07) & & 0.98 (2e-05) & 0.66 (2e-04)  \\ 
				\hline
				\multirow{2}*{ 0.3 } & 0.51 (2e-04) & 0.55 (2e-04) & 0.62 (2e-04) & 0.72 (2e-04) & 0.86 (1e-04) & 0.97 (2e-05) & 1 (7e-08) & \multirow{2}*{\diagbox[dir=SW]{~~~~~~}{~~~~~~}} & 0.97 (3e-05) & 0.42 (2e-04) & 0.13 (1e-04) & \multirow{2}*{ 0.7 } \\ 
				& 0.52 (2e-04) & 0.56 (3e-04) & 0.62 (2e-04) & 0.71 (2e-04) & 0.84 (1e-04) & 0.95 (4e-05) & 1 (6e-07) & & 0.99 (7e-06) & 0.83 (1e-04) & 0.57 (3e-04)  \\ 
				\hline
				\multirow{2}*{ 0.4 } & 0.52 (3e-04) & 0.59 (2e-04) & 0.7 (2e-04) & 0.84 (1e-04) & 0.97 (3e-05) & 1 (6e-08) & \multirow{2}*{\diagbox[dir=SW]{~~~~~~}{~~~~~~}} & 0.98 (2e-05) & 0.54 (3e-04) & 0.2 (2e-04) & 0.09 (8e-05) & \multirow{2}*{ 0.6 } \\ 
				& 0.52 (3e-04) & 0.59 (2e-04) & 0.68 (2e-04) & 0.82 (2e-04) & 0.95 (5e-05) & 1 (7e-07) & & 1 (2e-06) & 0.89 (9e-05) & 0.69 (2e-04) & 0.52 (2e-04)  \\ 
				\hline
				\multirow{2}*{ 0.5 } & 0.54 (2e-04) & 0.65 (2e-04) & 0.81 (1e-04) & 0.96 (4e-05) & 1 (2e-08) & \multirow{2}*{\diagbox[dir=SW]{~~~~~~}{~~~~~~}} & 0.99 (1e-05) & 0.61 (2e-04) & 0.25 (2e-04) & 0.12 (7e-05) & 0.06 (6e-05) & \multirow{2}*{ 0.5 } \\ 
				& 0.53 (2e-04) & 0.64 (3e-04) & 0.78 (2e-04) & 0.94 (5e-05) & 1 (1e-06) & & 1 (1e-06) & 0.92 (8e-05) & 0.75 (2e-04) & 0.6 (2e-04) & 0.49 (2e-04)  \\ 
				\hline
				\multirow{2}*{ 0.6 } & 0.56 (2e-04) & 0.74 (1e-04) & 0.93 (5e-05) & 1 (2e-07) & \multirow{2}*{\diagbox[dir=SW]{~~~~~~}{~~~~~~}} & 0.99 (8e-06) & 0.65 (3e-04) & 0.29 (2e-04) & 0.14 (1e-04) & 0.08 (7e-05) & 0.05 (5e-05) & \multirow{2}*{ 0.4 } \\ 
				& 0.55 (3e-04) & 0.72 (2e-04) & 0.91 (8e-05) & 1 (2e-06) & & 1 (9e-07) & 0.94 (7e-05) & 0.79 (2e-04) & 0.65 (3e-04) & 0.54 (2e-04) & 0.48 (2e-04)  \\ 
				\hline
				\multirow{2}*{ 0.7 } & 0.61 (2e-04) & 0.87 (1e-04) & 1 (1e-06) & \multirow{2}*{\diagbox[dir=SW]{~~~~~~}{~~~~~~}} & 0.99 (7e-06) & 0.67 (2e-04) & 0.31 (2e-04) & 0.15 (2e-04) & 0.09 (8e-05) & 0.06 (5e-05) & 0.05 (5e-05) & \multirow{2}*{ 0.3 } \\ 
				& 0.6 (2e-04) & 0.85 (1e-04) & 1 (5e-06) & & 1 (9e-07) & 0.95 (6e-05) & 0.81 (1e-04) & 0.67 (2e-04) & 0.58 (2e-04) & 0.51 (2e-04) & 0.46 (3e-04)  \\ 
				\hline
				\multirow{2}*{ 0.8 } & 0.69 (2e-04) & 0.99 (6e-06) & \multirow{2}*{\diagbox[dir=SW]{~~~~~~}{~~~~~~}} & 0.99 (5e-06) & 0.69 (2e-04) & 0.33 (2e-04) & 0.16 (1e-04) & 0.09 (1e-04) & 0.06 (6e-05) & 0.05 (5e-05) & 0.04 (4e-05) & \multirow{2}*{ 0.2 } \\ 
				& 0.69 (2e-04) & 0.98 (2e-05) & & 1 (7e-07) & 0.95 (4e-05) & 0.83 (1e-04) & 0.69 (2e-04) & 0.6 (2e-04) & 0.53 (2e-04) & 0.49 (2e-04) & 0.46 (2e-04)  \\ 
				\hline
				\multirow{2}*{ 0.9 } & 0.89 (1e-04) & \multirow{2}*{\diagbox[dir=SW]{~~~~~~}{~~~~~~}} & 0.99 (5e-06) & 0.69 (2e-04) & 0.33 (2e-04) & 0.17 (2e-04) & 0.1 (9e-05) & 0.07 (6e-05) & 0.05 (5e-05) & 0.05 (5e-05) & 0.04 (5e-05) & \multirow{2}*{ 0.1 } \\ 
				& 0.89 (1e-04) & & 1 (8e-07) & 0.95 (4e-05) & 0.83 (1e-04) & 0.71 (2e-04) & 0.62 (2e-04) & 0.55 (4e-04) & 0.51 (3e-04) & 0.48 (3e-04) & 0.46 (4e-04)  \\ 
				\hline
				& \diagbox[dir=SW]{~~~~~~}{~~~~~~} & & 0.85 & 0.75 & 0.65 & 0.55 & 0.45 & 0.35 & 0.25 & 0.15 & 0.05 & \backslashbox{$\eta$}{$\rho$} \\
				\hline
				\bottomrule[2pt]
				\multicolumn{3}{c}{} & \multicolumn{10}{c}{\Large Different parameters} \\
			\end{tabular}
			\label{graph_simul_same_diff_discrete_dist}
		}}
\end{table}
\begin{table}[!h]
	\centering
	\caption{Simulation results based on the frequency criterion.}
	\resizebox{5.7in}{1.6in}{
		\resizebox{\linewidth}{!}{
			\begin{tabular}{c|c|c|c|c|c|c|c|c|c|c|c|c}
				\multicolumn{10}{c}{\Large Same parameters} & \multicolumn{3}{c}{} \\
				\hline
				\bottomrule[2pt]
				\backslashbox{$\rho$}{$\eta$} & 0.05 & 0.15 & 0.25 & 0.35 & 0.45 & 0.55 & 0.65 & 0.75 & 0.85 & & \diagbox[dir=SW]{~~~~~~}{~~~~~~} & \\
				\hline
				\multirow{2}*{ 0.1 } & 0.51 (2e-04) & 0.52 (3e-04) & 0.55 (2e-04) & 0.58 (2e-04) & 0.65 (2e-04) & 0.76 (2e-04) & 0.91 (8e-05) & 0.99 (8e-06) & 1 (0e+00) & \multirow{2}*{\diagbox[dir=SW]{~~~~~~}{~~~~~~}} & 0.68 (2e-04) & \multirow{2}*{ 0.9 } \\ 
				& 0.51 (2e-04) & 0.53 (3e-04) & 0.56 (2e-04) & 0.6 (2e-04) & 0.66 (3e-04) & 0.75 (2e-04) & 0.87 (1e-04) & 0.97 (3e-05) & 1 (2e-08) & & 0.92 (9e-05)  \\ 
				\hline
				\multirow{2}*{ 0.2 } & 0.51 (2e-04) & 0.53 (3e-04) & 0.57 (3e-04) & 0.64 (3e-04) & 0.76 (1e-04) & 0.9 (9e-05) & 0.99 (8e-06) & 1 (0e+00) & \multirow{2}*{\diagbox[dir=SW]{~~~~~~}{~~~~~~}} & 0.96 (3e-05) & 0.16 (1e-04) & \multirow{2}*{ 0.8 } \\ 
				& 0.51 (3e-04) & 0.54 (2e-04) & 0.58 (2e-04) & 0.65 (2e-04) & 0.74 (2e-04) & 0.86 (2e-04) & 0.97 (3e-05) & 1 (3e-08) & & 0.99 (5e-06) & 0.72 (2e-04)  \\ 
				\hline
				\multirow{2}*{ 0.3 } & 0.51 (2e-04) & 0.55 (3e-04) & 0.62 (2e-04) & 0.74 (2e-04) & 0.89 (1e-04) & 0.99 (9e-06) & 1 (0e+00) & \multirow{2}*{\diagbox[dir=SW]{~~~~~~}{~~~~~~}} & 0.99 (6e-06) & 0.36 (3e-04) & 0.05 (4e-05) & \multirow{2}*{ 0.7 } \\ 
				& 0.52 (2e-04) & 0.56 (3e-04) & 0.62 (3e-04) & 0.72 (2e-04) & 0.85 (1e-04) & 0.97 (3e-05) & 1 (4e-08) & & 1 (5e-07) & 0.88 (2e-04) & 0.61 (3e-04)  \\ 
				\hline
				\multirow{2}*{ 0.4 } & 0.53 (3e-04) & 0.6 (2e-04) & 0.72 (2e-04) & 0.88 (8e-05) & 0.99 (1e-05) & 1 (0e+00) & \multirow{2}*{\diagbox[dir=SW]{~~~~~~}{~~~~~~}} & 1 (2e-06) & 0.49 (2e-04) & 0.08 (5e-05) & 0.02 (1e-05) & \multirow{2}*{ 0.6 } \\ 
				& 0.52 (3e-04) & 0.59 (3e-04) & 0.69 (2e-04) & 0.83 (1e-04) & 0.96 (4e-05) & 1 (6e-08) & & 1 (1e-07) & 0.92 (7e-05) & 0.73 (2e-04) & 0.54 (3e-04)  \\ 
				\hline
				\multirow{2}*{ 0.5 } & 0.54 (2e-04) & 0.67 (2e-04) & 0.85 (1e-04) & 0.99 (1e-05) & 1 (0e+00) & \multirow{2}*{\diagbox[dir=SW]{~~~~~~}{~~~~~~}} & 1 (2e-06) & 0.56 (3e-04) & 0.11 (1e-04) & 0.02 (2e-05) & 0.01 (7e-06) & \multirow{2}*{ 0.5 } \\ 
				& 0.54 (3e-04) & 0.65 (2e-04) & 0.8 (2e-04) & 0.95 (4e-05) & 1 (5e-08) & & 1 (7e-08) & 0.95 (5e-05) & 0.79 (2e-04) & 0.62 (2e-04) & 0.5 (2e-04)  \\ 
				\hline
				\multirow{2}*{ 0.6 } & 0.57 (2e-04) & 0.79 (2e-04) & 0.97 (3e-05) & 1 (2e-08) & \multirow{2}*{\diagbox[dir=SW]{~~~~~~}{~~~~~~}} & 1 (1e-06) & 0.59 (3e-04) & 0.13 (9e-05) & 0.03 (2e-05) & 0.01 (1e-05) & 0.01 (5e-06) & \multirow{2}*{ 0.4 } \\ 
				& 0.56 (3e-04) & 0.74 (2e-04) & 0.93 (6e-05) & 1 (1e-07) & & 1 (6e-08) & 0.96 (6e-05) & 0.82 (1e-04) & 0.67 (3e-04) & 0.56 (2e-04) & 0.49 (3e-04)  \\ 
				\hline
				\multirow{2}*{ 0.7 } & 0.64 (3e-04) & 0.93 (6e-05) & 1 (4e-08) & \multirow{2}*{\diagbox[dir=SW]{~~~~~~}{~~~~~~}} & 1 (8e-07) & 0.62 (2e-04) & 0.14 (1e-04) & 0.03 (2e-05) & 0.01 (9e-06) & 0.01 (5e-06) & 0 (4e-06) & \multirow{2}*{ 0.3 } \\ 
				& 0.61 (2e-04) & 0.88 (1e-04) & 1 (5e-07) & & 1 (5e-08) & 0.96 (3e-05) & 0.84 (1e-04) & 0.69 (2e-04) & 0.59 (2e-04) & 0.52 (3e-04) & 0.47 (3e-04)  \\ 
				\hline
				\multirow{2}*{ 0.8 } & 0.75 (2e-04) & 1 (5e-07) & \multirow{2}*{\diagbox[dir=SW]{~~~~~~}{~~~~~~}} & 1 (9e-07) & 0.63 (2e-04) & 0.15 (1e-04) & 0.03 (3e-05) & 0.01 (1e-05) & 0.01 (5e-06) & 0 (4e-06) & 0 (4e-06) & \multirow{2}*{ 0.2 } \\ 
				& 0.71 (3e-04) & 1 (5e-06) & & 1 (2e-08) & 0.97 (3e-05) & 0.85 (1e-04) & 0.71 (2e-04) & 0.61 (2e-04) & 0.54 (2e-04) & 0.5 (3e-04) & 0.47 (3e-04)  \\ 
				\hline
				\multirow{2}*{ 0.9 } & 0.94 (6e-05) & \multirow{2}*{\diagbox[dir=SW]{~~~~~~}{~~~~~~}} & 1 (9e-07) & 0.63 (2e-04) & 0.15 (1e-04) & 0.04 (3e-05) & 0.01 (1e-05) & 0.01 (8e-06) & 0 (4e-06) & 0 (4e-06) & 0 (3e-06) & \multirow{2}*{ 0.1 } \\ 
				& 0.92 (7e-05) & & 1 (3e-08) & 0.97 (3e-05) & 0.85 (1e-04) & 0.73 (2e-04) & 0.63 (2e-04) & 0.56 (3e-04) & 0.52 (3e-04) & 0.49 (2e-04) & 0.46 (3e-04)  \\ 
				\hline
				& \diagbox[dir=SW]{~~~~~~}{~~~~~~} & & 0.85 & 0.75 & 0.65 & 0.55 & 0.45 & 0.35 & 0.25 & 0.15 & 0.05 & \backslashbox{$\eta$}{$\rho$} \\
				\hline
				\bottomrule[2pt]
				\multicolumn{3}{c}{} & \multicolumn{10}{c}{\Large Different parameters} \\
			\end{tabular}
			\label{graph_simul_same_diff_discrete_freq}
		}}
\end{table}

\section{Real Data Analysis}
\label{sec:real}
\begin{figure}[H]
	\centering
	\includegraphics[scale=0.6]{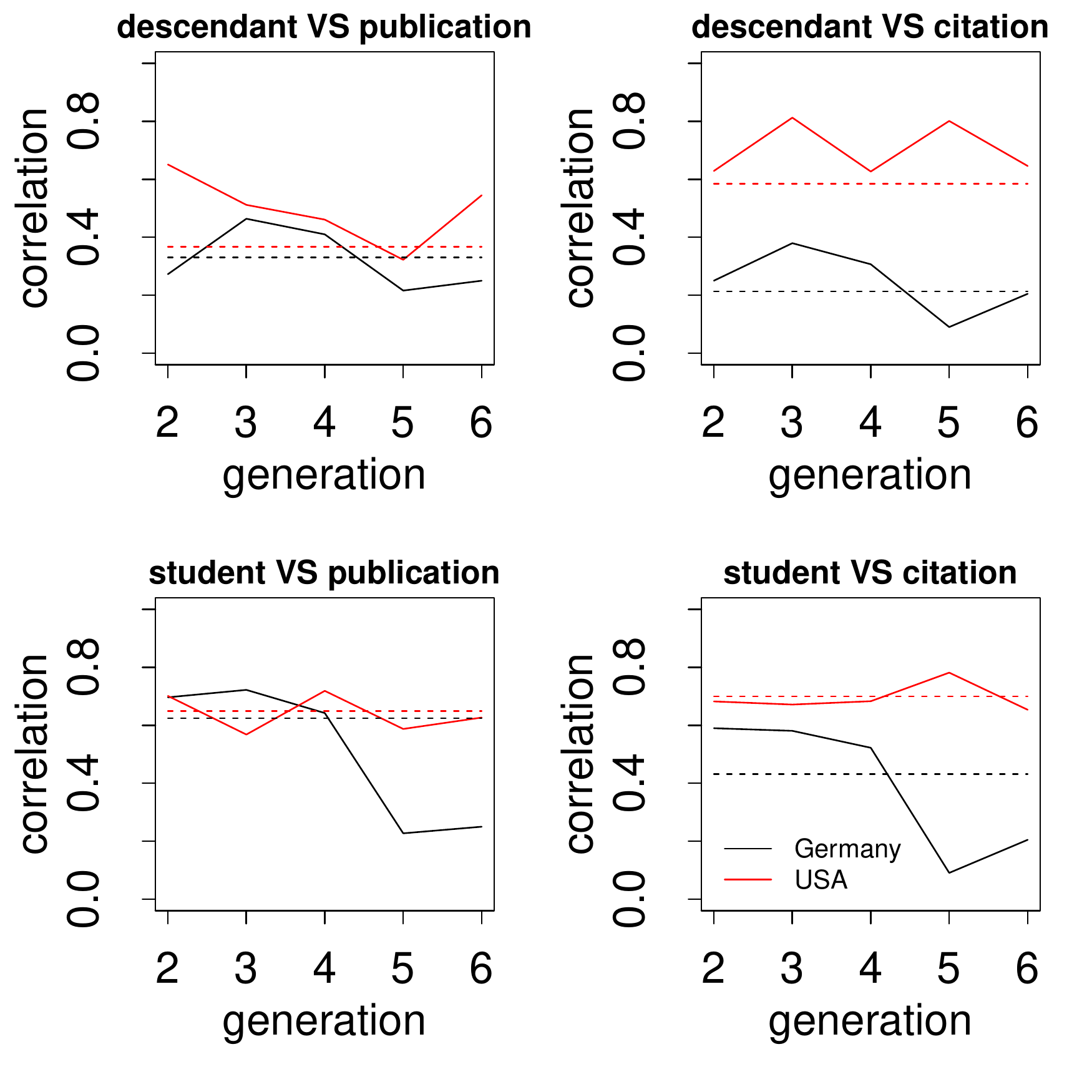}
	\caption{The sample Pearson correlations of each generation on real data. Each sub-figure corresponds to one kind of correlations. The solid line is the sample Pearson correlations calculated generation by generation, while the dotted line represents the sample Pearson correlation calculated by putting all data into a line without changing the paired relationship.}
	\label{cor_real}
\end{figure}

Our real data analysis is applied on a dataset that is integrated from two public online databases, the Mathematics Genealogy Project (MGP) \citep{horowitz2008mathematicians} and MathSciNet \citep{richert2011authors}. MGP, a service provided by North Dakota State University, contains information about PhDs in mathematic discipline, including their advisors, dissertation titles, graduate colleges, graduate nations, graduate years, the numbers of their PhD students (stud) and the numbers of their PhD descendants (desc). Meanwhile, MathSciNet in the Mathematical Reviews (MR) Database contains information on the numbers of total publications (publ) and total citations (cita) for each person in MGP. After excluding the cases of co-advisors, every scholar can be involved into one tree structure according to advisor-PhD relationships. Our target is to investigate advisor-PhD partnerships in mathematical area across countries. Two genealogy trees with 6 generations are selected out and some summary information is displayed in Table~\ref{tree_infor}. As we can see, their main difference is graduate nation, i.e., most of PhDs within two trees are graduated from Germany and USA respectively. To some extent, the cooperative relationship between advisors and PhDs can be reflected by the relationship between student enrollment and academic achievements of advisors. Thus, the following four kinds of correlations are considered: $\text{Cor}_i(\text{desc}, \text{publ})$, $\text{Cor}_i(\text{desc}, \text{cita})$, $\text{Cor}_i(\text{stud}, \text{publ})$, $\text{Cor}_i(\text{stud}, \text{cita})$, $i = 1, 2$. Figure~\ref{cor_real} displays the sample correlations of each generations on real data. We can see that the correlations generally decline or mildly fluctuate as the generation growing and the red solid line is almost always above the black one which satisfy the assumptions that $f(i; \rho)$ is a decreasing function of generation $i$ and an increasing function of $\rho$, respectively, proposed in Section~\ref{sec:model}.
\begin{table}[!h]
	\vspace{-0.8cm}
	\centering
	\caption{Summary information about genealogy trees. `Year of Root' represents the PhD graduate year of the root node. `Number of Records' is the total number of persons within the corresponding tree. `Proportion of the Country' shows the proportion of nodes whose PhD graduate nations are the same to the 'Country' item.}
	\begin{tabular}{c|c|c}
		Attribute & Tree 1 & Tree 2 \\
		\midrule[2pt]
		Country & Germany & USA \\
		\hline
		Year of Root & 1935 & 1935 \\
		\hline
		Numer of Records & 1257 & 887 \\
		\hline
		Proportion of the Country & 86.16\% & 67.01\% \\
	\end{tabular}
	\label{tree_infor}
\end{table}

There is one thing worth mentioning, given that there is no significant linear relationship between data of adjacent generations, such as the number of students of advisors and that of the corresponding PhDs. Thus, the raw data are directly treated as incremental data in subsequent analysis. Before real data analysis, mimic datasets are synthesized to validate our method: first, fit real data to DSPGM to obtain MLE of parameters; then, set these estimators as true parameters, and randomly generate a mimic dataset with the same tree topology as real one. For each kind of correlation in Figure~\ref{cor_real}, $TD \Delta \theta$ algorithm is applied as follows:
\begin{itemize}
	\setlength{\itemsep}{0pt}
	\setlength{\parsep}{0pt}
	\setlength{\parskip}{0pt}
	\item[Step 1:] Apply $TD \Delta \theta$ algorithm to calculate two included angles $\Delta \theta_{Germany}$, $\Delta \theta_{USA}$;
	\item[Step 2:] Repeat Step 1,2,3 for 1000 times and calculate the proportion of $\Delta \theta_{Germany} > \Delta \theta_{USA}$;
	\item[Step 3:] Repeat Step 4 for 100 times and get the mean and standard deviation of the proportion.
\end{itemize}
For comparison, we also transform each mimic dataset into a vector and then calculate the Pearson correlation coefficient. The results are shown in Table~\ref{mimic_graph}. After then, two methods are applied on real dataset and the results are displayed in Table~\ref{real_graph}.

\begin{table}[H]
	\centering
	\caption{Comparison of results on mimic data.}
	\begin{tabular}{l||l l}
		\multirow{2}* & \multicolumn{2}{c}{Proportion} \\
		\cline{2-3}
		& $\Delta \theta_{Germany} > \Delta \theta_{USA}$ & $\rho_{Germany} < \rho_{USA}$ \\
		\hline
		\multirow{1}*{descendant VS publication} & 81.03\% (0.011) & 71.15\% (0.017) \\
		\hline
		\multirow{1}*{descendant VS citation} & 99.98\% (0.001) & 100\% (0) \\
		\hline
		\multirow{1}*{student VS publication} & 91.99\% (0.009) & 76.98\% (0.014) \\
		\hline
		\multirow{1}*{student VS citation} & 99.85\% (0.001) & 100\% (0) \\
	\end{tabular}
	\label{mimic_graph}
\end{table}
\begin{table}[H]
	\vspace{-0.8cm}
	\centering
	\caption{Comparison of results on real data.}
	\begin{tabular}{l||l |l}
		& $\Delta \theta_{Germany} \ \ \Delta \theta_{USA}$ & $\rho_{Germany} \ \ \rho_{USA}$ \\
		\hline
		\multirow{1}*{descendant VS publication} & $32.79^{\circ} > 29.09^{\circ}$ & $0.33 < 0.366$ \\
		\hline
		\multirow{1}*{descendant VS citation} & $33.78^{\circ} > 21.71^{\circ}$ & $0.213 < 0.585$ \\
		\hline
		\multirow{1}*{student VS publication} & $31.26^{\circ} > 28.51^{\circ}$ & $0.624 < 0.649$ \\
		\hline
		\multirow{1}*{student VS citation} & $32.59^{\circ} > 24.33^{\circ}$ & $0.432 < 0.7$ \\
	\end{tabular}
	\label{real_graph}
	\vspace{-0.6cm}
\end{table}

In both simulation and real data analysis, we can find that the calculated $\Delta \theta$s of USA are almost uniformly smaller than $\Delta \theta$s of Germany, which indicates that the academic cooperation between PhD graduates and their advisors in USA is more close than that of Germany. The reasons may be concluded in two aspects: on one hand, in the German higher education system, doctoral education can be regarded as a pretty unsystematic educational path or an unstructured approach and proposed that the German doctoral student is `free as a bird' \citep{roebken2007postgraduate}, while, doctoral education in USA has a relatively complete structure, and the doctoral admissions and training are all managed by the graduate school, so that there will be no lack of quality assurance and proper supervision \citep{walker2008doctoral}; on the other hand, there is a higher proportion of PhD graduates major in science eventually go into full-time academic positions in USA than that in Germany \citep{cyranoski2011education}. If PhD graduates continue to do research, they are more likely to maintain academic cooperation with their supervisors, so a high proportion of graduates doing research may reflect a closer relationship between supervisors and PhD graduates. Both of these factors reflect that the academic cooperation between PhD graduates and advisors is closer in USA than that in Germany, which is consistent with our findings.

Although the same results can also be obtained by the Pearson correlation method, but in simulation, it can be easily found that $TD \Delta \theta$ achieves a higher accuracy than the traditional Pearson method.

\section{Discussion and Conclusion}
\label{sec:conclusion}

In this paper, a geometric statistic is proposed to display the geometric interpretation of the tree correlation. Both simulation studies and real data analysis show that our method can demonstrate the degree of tree correlation qualitatively with $TD \Delta \theta$ algorithm. Moreover, the reliability of $\Delta \theta$ is theoretically proved through quantile ellipse.

\begin{figure}
	\vspace{-0.8cm}
	\centering
	\includegraphics[scale=0.21]{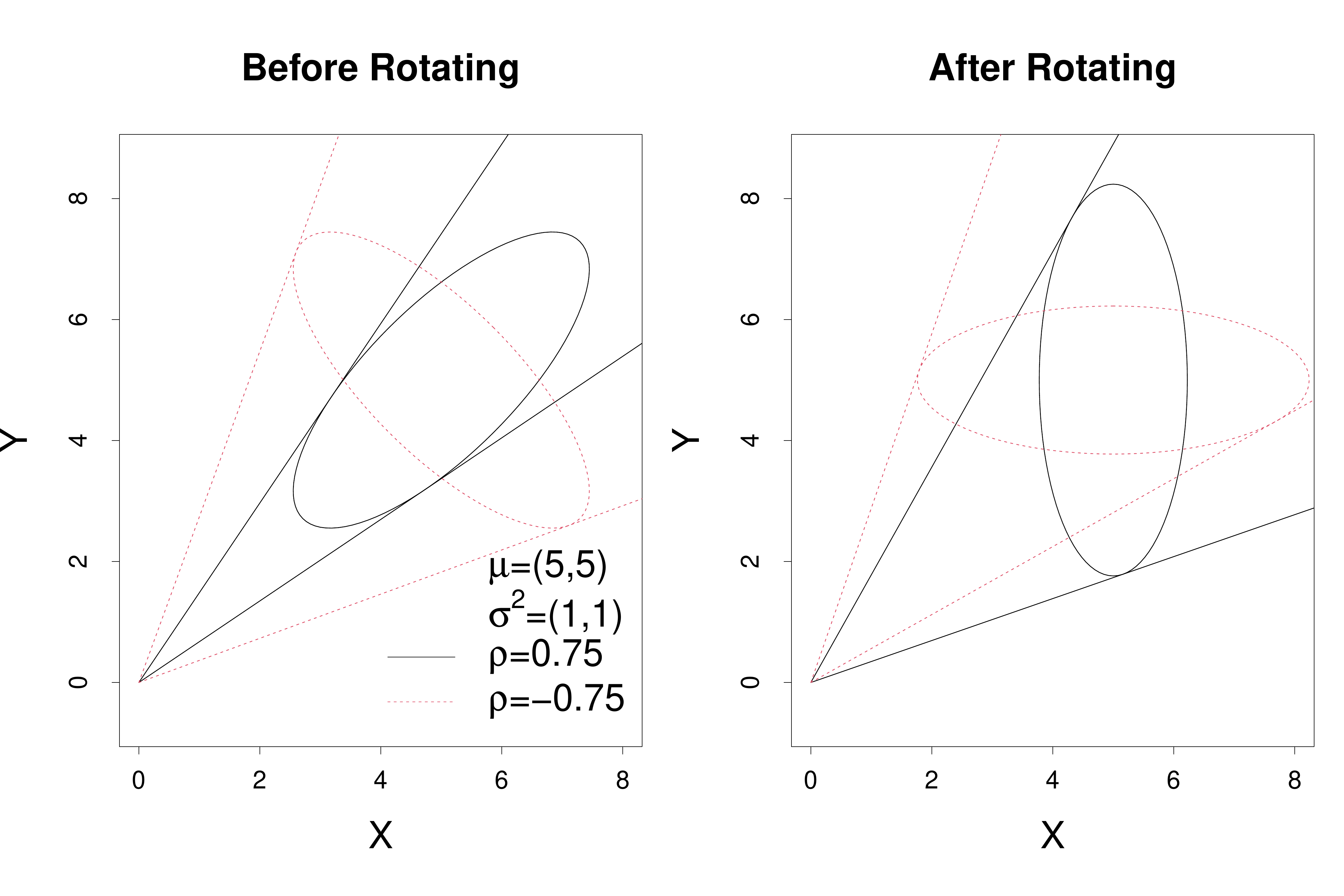}
	\caption{Quantile ellipses for bivariate Gaussian distributions before and after rotating. In each sub-figure, the two quantile ellipses are corresponding two bivariate Gaussian distributions with the same parameters, except for $\rho$, which are inverse numbers of each other. The ellipses in the right sub-figure is obtained by rotating the ellipses in the left sub-figure $45^{\circ}$ counterclockwise with $\mu=(5,5)$ as the center.}
	\label{rho1_1}
\end{figure}

Note that we assume $f(i;\rho)$ is a positive number in the current model because we mainly concern the positive correlation damping pattern. Extending the $TD \Delta \theta$ algorithm without this constraint is straightforward. Actually, for bivariate Gaussian with fixed expectation and variance, it's not hard to prove that its $\Delta \theta$ is a decreasing function of $\rho$ in domain $[-1, 1]$ with formula~\eqref{anglerho}. Thus, even if $\rho < 0$, all conclusions in the paper still stand, as long as we adopt $(-\Delta \boldsymbol{X}, \Delta \boldsymbol{Y})$ or $(\Delta \boldsymbol{X}, -\Delta \boldsymbol{Y})$ to get $\Delta \hat{\theta}$. On the other hand, we also find, for two bivariate Gaussian distributions with the same parameters, except for $\rho$, which are inverse numbers of each other, if their quantile ellipses are rotated $45^{\circ}$ counterclockwise around the center, their corresponding included angles become the same, just as shown in Figure~\ref{rho1_1}. This conclusion means that, when $f(i;\rho)$ is negative, all of these properties proposed in the paper are still validated for the corresponding angle after rotating, and moreover, this angle can be used to compare the relative magnitude of the absolute value of tree correlation, even though the direction of this correlation is unknown.

In addition, the proposed $\Delta \theta$ has strong applicability since it can be extended to other models. In detail, it is unnecessary to model the data via increments, only if independent distributions could be obtained from the raw data after some preprocessing. For example, if the original tree-shaped dataset, $\{ (X^{i,j}_t, Y^{i,j}_t) \}$, is modeled by ratio as follows:
\begin{equation}
\left(\begin{array}{c}
X^{i,j}_t/X^{i,j}_{t-1} \\
Y^{i,j}_t/Y^{i,j}_{t-1}
\end{array} \right)
\sim N
\begin{pmatrix}
\begin{pmatrix}
m^{i,j} \\ n^{i,j}
\end{pmatrix}
&
,
&
\begin{pmatrix}
\sigma_1^2 & f(i; \rho) \sigma_1 \sigma_2 \\
f(i; \rho) \sigma_1 \sigma_2 & \sigma_2^2
\end{pmatrix}
\end{pmatrix}.
\end{equation}
Then, we can construct a new set of random variables $\{ (X^{i,j}_{t\#}, Y^{i,j}_{t\#}) \}$, which satisfy
\begin{equation*}
\begin{pmatrix}
X^{1,1}_{0\#} \\ Y^{1,1}_{0\#}
\end{pmatrix}
=
\begin{pmatrix}
0 \\ 0
\end{pmatrix}
,
\begin{pmatrix}
X^{i,j}_{t\#} \\ Y^{i,j}_{t\#}
\end{pmatrix}
=
\begin{pmatrix}
X^{i,j}_t/X^{i,j}_{t-1} + X^{i,j}_{(t-1)\#} \\ Y^{i,j}_t/Y^{i,j}_{t-1} + X^{i,j}_{(t-1)\#}
\end{pmatrix} \ ,\quad \text{for} \ i =1, \cdots.
\end{equation*}
Finally, the data can be transformed to the following form as before:
\begin{equation}
\left(\begin{array}{c}
X^{i,j}_{t\#} - X^{i,j}_{(t-1)\#} \\
Y^{i,j}_{t\#} - Y^{i,j}_{(t-1)\#}
\end{array} \right)
\sim N
\begin{pmatrix}
\begin{pmatrix}
m^{i,j} \\ n^{i,j}
\end{pmatrix}
&
,
&
\begin{pmatrix}
\sigma_1^2 & f(i; \rho) \sigma_1 \sigma_2 \\
f(i; \rho) \sigma_1 \sigma_2 & \sigma_2^2
\end{pmatrix}
\end{pmatrix}.
\end{equation}

Even so, there are still some problems to be solved. For instance, the statistical properties of $\Delta \theta$ have not been fully explored, such as its real or approximate null distribution; the setting of parameter $\mu_i^*$ is also artificial and can not be determined in an adaptive manner based on practical situation if $f(i;\rho)$ is unknown. Moreover, in the normalization procedure, those linear transformations are conducted based on the parameter estimates, instead of their true values. The measurement of the error caused by this operation for the estimation of $\Delta \theta$ needs further study. These will be the focus of our future work.

\bibliographystyle{Chicago}
\bibliography{Bibliography-MM-MC}

\end{document}